\documentclass[unnumsec,webpdf,modern,large]{oup-authoring-template}
\usepackage{caption}
\usepackage{float}
\usepackage{booktabs}
\usepackage{rotating}
\usepackage{pgffor}
\usepackage{tikz}
\graphicspath{{figs/}}

\usepackage{seqsplit}
\newcommand{\dna}[1]{\texttt{\seqsplit{#1}}}

\newcommand{\siref}[1]{\ref{#1}}
\newcommand{\siautoref}[1]{\autoref{#1}}

\providecommand{\Endparasplit}{}

\newif\ifmergernapreprint
\mergernapreprinttrue
\newcommand{\mergernaMonthName}[1]{%
	\ifcase#1\or January\or February\or March\or April\or May\or June\or July\or August\or September\or October\or November\or December\fi}
\newcommand{\preprintdatestamp}{\mergernaMonthName{\month}~\number\day,~\number\year}
\input{preprint-config}

\theoremstyle{thmstyleone}%

\theoremstyle{thmstyletwo}%
\theoremstyle{thmstylethree}%

\begin{document}

\ifmergernapreprint
\journaltitle{}
\DOI{}
\copyrightyear{}
\pubyear{\the\year}
\access{Preprint compiled on \preprintdatestamp\christmastree}
\appnotes{Preprint}
\else
\journaltitle{Journal Title Here}
\DOI{DOI HERE}
\copyrightyear{2022}
\pubyear{2025}
\access{Advance Access Publication Date: Day Month Year}
\appnotes{Paper}
\fi

\firstpage{1}

\title[MERGE-RNA]{MERGE-RNA: a physics-based model to predict RNA secondary structure ensembles with chemical probing}

\author[1]{Giuseppe Sacco\ORCID{0009-0001-0594-8349}}
\author[2]{Jianhui Li}
\author[2]{Redmond P. Smyth\ORCID{0000-0002-1580-0671}}
\author[1]{Guido Sanguinetti\ORCID{0000-0002-6663-8336}}
\author[1,$\ast$]{Giovanni Bussi\ORCID{0000-0001-9216-5782}}

\authormark{Sacco et al.}

\address[1]{\orgname{Scuola Internazionale Superiore di Studi Avanzati}, \orgaddress{\street{via Bonomea 265}, \postcode{34136}, \country{Italy}}}
\address[2]{\orgdiv{Architecture et Réactivité de l'ARN}, \orgname{Université de Strasbourg, CNRS, Institute of Molecular and Cellular Biology (IBMC)}, \orgaddress{\street{2 Allée Konrad Roentgen}, \postcode{67000}, \state{Strasbourg}, \country{France}}}

\corresp[$\ast$]{Corresponding author. \href{mailto:bussi@sissa.it}{bussi@sissa.it}}

\received{Date}{0}{Year}
\revised{Date}{0}{Year}
\accepted{Date}{0}{Year}

\abstract{RNA function is tied to secondary structure, operating through dynamic and heterogeneous structural ensembles. While current analysis tools typically output single static structures or averaged contact maps, chemical probing methods like DMS capture nucleotide-resolution signals representing the full structural ensemble, which remain difficult to interpret structurally. To address this, we present MERGE-RNA, a framework that describes and outputs RNA as a structural ensemble. By modeling the physics of the experimental pipeline, MERGE-RNA learns a small set of transferable and interpretable parameters, enabling the integration of measurements across different molecules, probe concentrations, and replicates in a single optimization to improve robustness. Our model employs a maximum-entropy principle to predict thermodynamic populations, with the minimal adjustments necessary to align the ensemble with experimental data. We validate MERGE-RNA on diverse RNAs, showing that it achieves structural accuracy surpassing standard pseudo-free-energy methods and yields ensembles better recapitulating measured DMS reactivity. Applied to the \textit{V. vulnificus} adenine riboswitch, MERGE-RNA recovers the NMR-resolved conformations and their ligand-induced rearrangement, with population shifts matching the NMR-derived $K_d$. In a designed RNA construct for which we report new DMS data, MERGE-RNA deconvolves mixed states to reveal transient intermediate populations involved in strand displacement, dynamics invisible to methods based on enumerating a small number of structures.}


\maketitle
\section*{Graphical Abstract}
\begin{figure}[h]
\centering
\includegraphics[width=.66\textwidth]{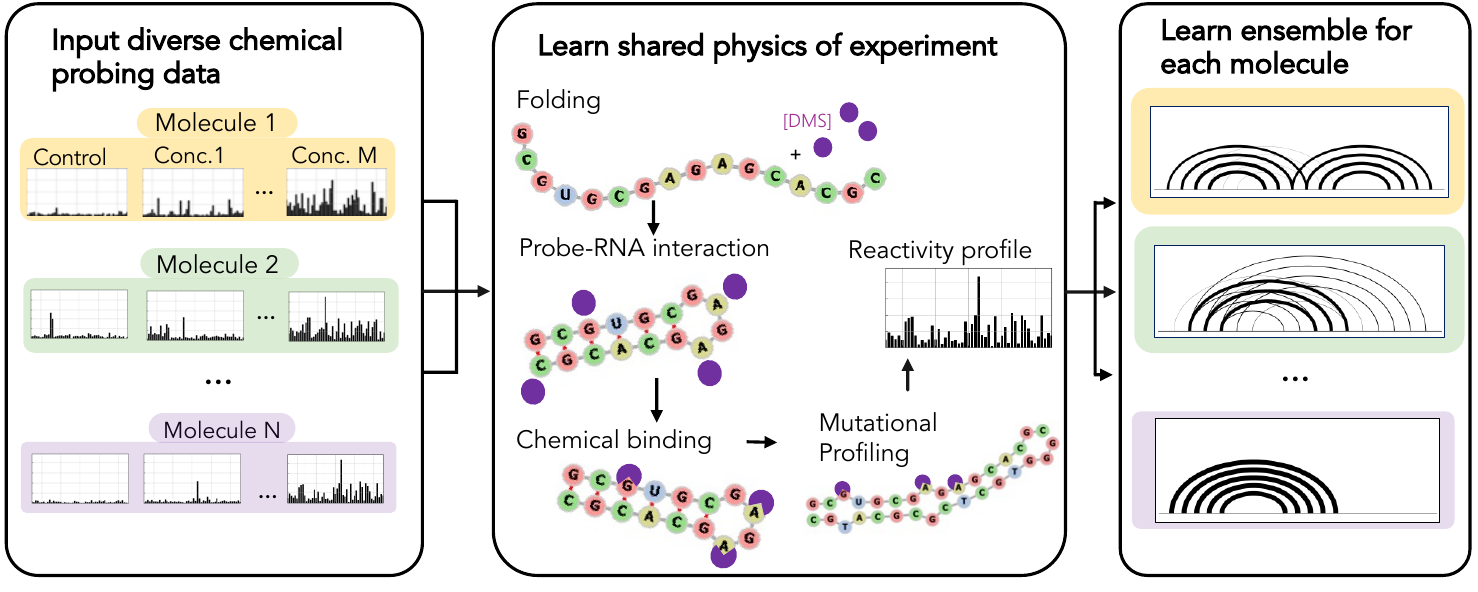}
\label{fig:graphical_abstract}
\end{figure}

\twocolumn
\section*{Introduction}
\label{sec:intro}

RNA molecules play central roles in cellular processes, ranging from catalysis (ribozymes and ribosomes) \cite{doudnaChemicalRepertoireNatural2002,ramakrishnanRibosomeStructureMechanism2002} to gene regulation (including riboswitches and microRNAs) \cite{morrisRiseRegulatoryRNA2014}.
This functional versatility arises from their ability to fold into specific secondary and tertiary structures \cite{butcherMolecularInteractionsThat2011}.
Importantly, RNA molecules exist as thermodynamic ensembles at both the secondary \cite{boseCausesFunctionsTherapeutic2024} and tertiary \cite{kenRNAConformationalPropensities2023} levels, with function arising from stable conformations, minority states, or dynamic equilibria.
While tertiary structure modeling is computationally demanding \cite{buRNAPuzzlesBlindPredictions2025}, secondary structure thermodynamic models allow efficient computation of optimal structures \cite{nussinovFastAlgorithmPredicting1980} and base-pairing probabilities \cite{mccaskillEquilibriumPartitionFunction1990}.
However, thermodynamic models face significant limitations from fundamental approximations:
the nearest-neighbor approximation \cite{tinocoImprovedEstimationSecondary1973},
the neglect of pseudoknots \cite{nussinovFastAlgorithmPredicting1980}, and
the hierarchical folding assumption \cite{tinocoHowRNAFolds1999} may make it difficult to fully capture long-range interactions and context-dependent effects.
To address these limitations, researchers have integrated co-evolutionary analysis or chemical probing data.
Dimethyl sulfate (DMS) \cite{zaugAnalysisStructureTetrahymena1995,kubotaProgressChallengesChemical2015} has become a powerful tool for structural analysis, providing single-nucleotide resolution data both \textit{in vitro} and \textit{in vivo}.
DMS reacts preferentially with unpaired adenines and cytosines, with modifications detected as mutations during sequencing (mutational profiling, MaP) \cite{kubotaProgressChallengesChemical2015}.
However, interpreting this data is challenging: measured reactivities report on structural ensembles rather than single conformations, and experimental processes introduce systematic biases from reverse transcription and sequencing artifacts.
Current approaches typically convert measured reactivities into pseudo-free-energy restraints \cite{deiganAccurateSHAPEdirectedRNA2009}, but this modifies predicted ensembles even when baseline models already agree with data \cite{washietlRNAFoldingSoft2012}, and the coefficients used to compute pseudo-free-energies are \Endparasplit
based on heuristics rather than physical principles.
\begin{figure*}[t]
    \centering
    \includegraphics[width=\textwidth, trim={0 60pt 0 85pt}, clip]{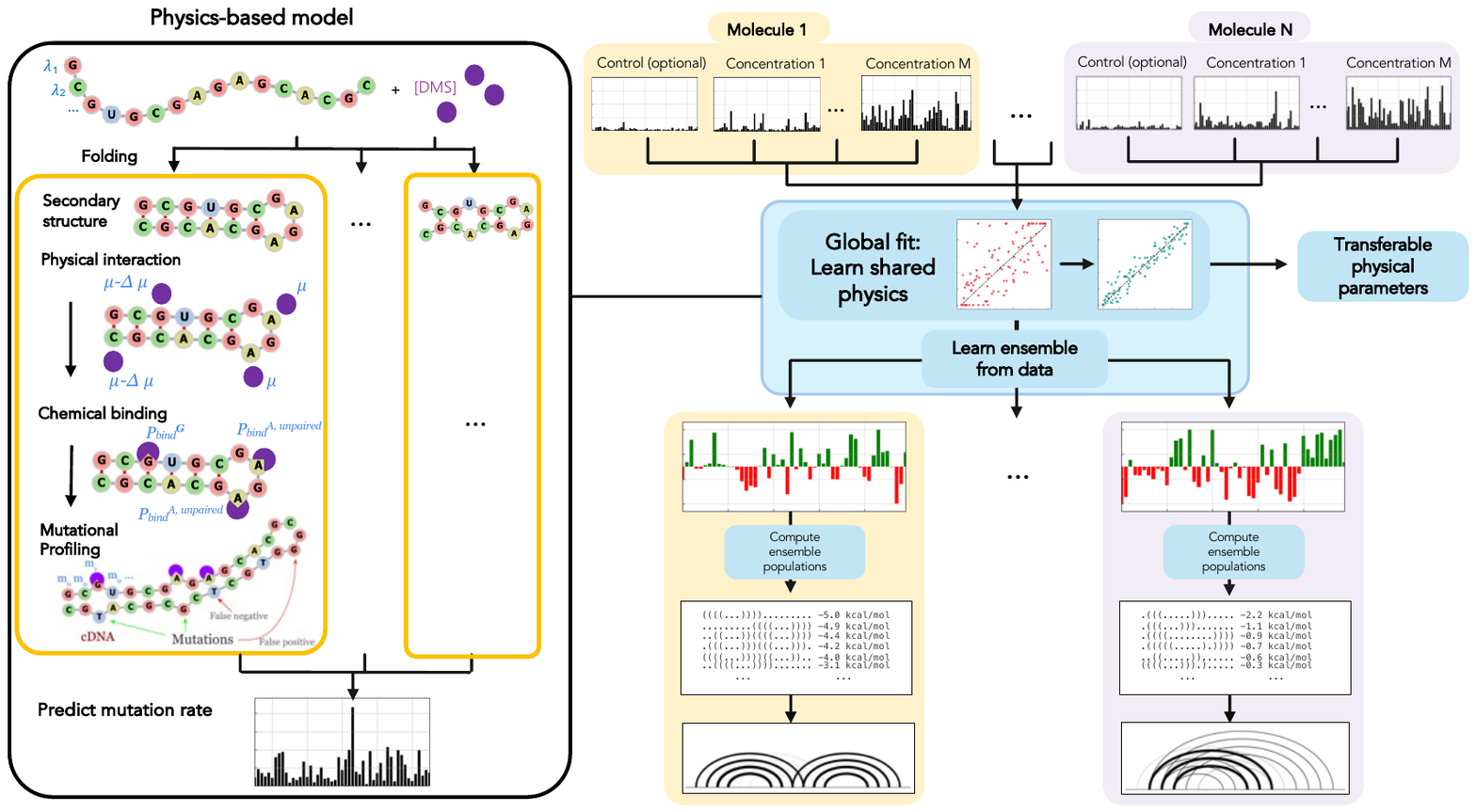}
    \caption{
    Schematic overview of the method.
    Our approach builds a physical model---illustrated on the left side of the figure---that represents the entire pipeline of chemical probing experiments. This includes RNA folding into an ensemble of secondary structures, probe binding, adduct formation, and the final mutational profiling readout.
    The model is trained on chemical probing data from multiple RNA sequences, probe concentrations, and experimental replicates. From these data, it learns a set of physically meaningful parameters shared across all experiments ($\mu_r$, $\Delta\mu_{\text{pairing}}$,  $p_\mathrm{bind}(A)$, $p_\mathrm{bind}(C)$, $p_\mathrm{bind}(G)$, $p_\mathrm{bind}(U)$, $m_0$ and $m_1$, as defined in the main text). In addition, it estimates sequence-specific soft constraints $\lambda_i$ (one per nucleotide position).
    Once trained, the model can predict the structural ensemble of each RNA at any probe concentration. Importantly, it can also extrapolate to zero probe concentration, providing the unperturbed structural ensemble.
    }
    \label{fig:1}
\end{figure*}
Despite widespread practice, focusing on a single minimum free-energy (MFE) structure can be misleading as RNA molecules are known to adopt an ensemble of conformations, of which the MFE structure is often only a minor component.
In fact, ignoring structural heterogeneity can pose problems both for interpreting RNA function and for properly modeling chemical probing experiments.

In this work we present MERGE-RNA (Multi-system Ensemble Refinement via Generalizable parameters Estimation), a unified physics-based model that describes chemical probing experiments with physically meaningful parameters and predicts RNA secondary structure ensembles.
Our approach combines a standard thermodynamic folding model with explicit probe-RNA interactions characterized by two key parameters: a chemical potential ($\mu_r$) setting the probe concentration scale, and an energetic penalty ($\Delta\mu_{\text{pairing}}$) for probe binding to paired nucleotides.
We employ maximum entropy inference to determine sequence-specific soft constraints ($\lambda_{i}$) representing minimal adjustments for model-data consistency, avoiding artificial constraints when baseline predictions already match data.
The model explicitly accounts for mutational profiling artifacts, including false positives, false negatives, and position-specific biases.
Crucially, our framework enables the simultaneous analysis of data from multiple RNA sequences, probe concentrations, and experimental replicates within a unified model, leveraging the shared physical parameters to improve robustness and extract transferable experimental knowledge.

After outlining the model and inference procedure, we benchmark it on a set of well-characterized structured RNAs for which DMS probing data are available across multiple concentrations and replicates under consistent experimental conditions \citep{bohnNanoDMSMaPAllowsIsoformspecific2023}. This homogeneity makes the dataset well suited to assess both the transferability of the inferred physical parameters and the structural accuracy of the predicted ensembles, which we benchmark against the standard pseudo-free energy approach \citep{deiganAccurateSHAPEdirectedRNA2009}.
On these systems, we also show that no single static structure fully explains the probing data, indicating that conformational heterogeneity is required even for RNAs traditionally treated as well-folded, and that the ensembles predicted by MERGE-RNA significantly improve the agreement with probing data when compared to the traditional pseudo-free energy approach.
For a more stringent validation of the inferred heterogeneity, we then turn to the \textit{V. vulnificus} adenine riboswitch, for which NMR provides independent ground truth on the co-existing conformations and their ligand-dependent populations \citep{reiningThreestateMechanismCouples2013} and which has benchmarked other deconvolution methods \citep{morandiGenomescaleDeconvolutionRNA2021,olsonDiscoveryLargescaleCellstateresponsive2022}: MERGE-RNA recovers the structural details of each NMR-resolved conformer (except the pseudoknot, excluded by the thermodynamic baseline) and reproduces the ligand-dependent population shifts, matching the NMR-derived $K_d$.
Finally, we demonstrate accurate deconvolution of mixed states in synthetic bistable constructs, and leverage experimental data to reveal intermediate, suboptimal populations involved in dynamic processes (e.g., strand displacement) that are missed by traditional approaches.

\section*{Methods}\label{sec:methods}

\subsection*{Physical Model}\label{subsec:physical_model}

Our model follows the experimental pipeline of chemical probing experiments, as illustrated in \autoref{fig:1}.
The underlying physics remains consistent across experiments using the same reagent, probing time, and temperature, allowing us to simultaneously fit multiple experimental datasets within a unified framework.
By maximizing the likelihood of the model to reproduce the observed mutation rates,
our approach estimates physically meaningful parameters (detailed below) and identifies the ensemble of secondary structures most consistent with both the underlying physics and experimental evidence.

In this section we aim at introducing the physical model underlying our approach, while in the Supplementary Information we provide a more in-depth description and rigorous discussion.

\subsubsection*{RNA folding and chemical probing}\label{subsubsec:rna_folding}

\paragraph*{\textit{Folding}}

To model the RNA secondary structure ensemble, we utilize as baseline the ViennaRNA package \citep{lorenzViennaRNAPackage202011}. The probability to observe a given secondary structure $s$ at temperature $T$ is
defined as
\begin{align}
    \mathbb{P}(s) = \frac{\exp\left(-\frac{F(s)}{\mathrm{k_B}T}\right)}{z}
        \label{eq:ps_maintext}
\end{align}
where $z$ is the partition function, $\mathrm{k_B}$ is the Boltzmann constant, and the baseline $F(s)=F_0(s)$ is
computed by the underlying ViennaRNA thermodynamic model, including the standard nearest-neighbour contributions (e.g. stacking interactions, hairpin, internal, bulge and multiloop terms).
To improve predictions, we refine the ensemble to align with experimental data while minimally perturbing the baseline model.
Using the principle of maximum entropy \citep{demartinoIntroductionMaximumEntropy2018}, we optimize
sequence-specific parameters ($\lambda_i$) that act as soft constraints for each nucleotide position \cite{washietlRNAFoldingSoft2012}.
These soft constraints modify the baseline ViennaRNA-predicted energy of pairing and remain consistent across experimental replicates and probe concentrations for a given RNA sequence.
This results in a structure-dependent energy correction
\begin{equation}
    \Delta F_{\text{opt}}(s) = \sum_{i \in \text{paired}(s)} \lambda_i
\end{equation}
where $\text{paired}(s)$ denotes the list of paired nucleobases in structure $s$.

\paragraph*{\textit{Probe-RNA interaction}}
We separately model the probe reactivity in two steps \cite{calonaciMolecularDynamicsSimulations2023}: a reversible physical binding, described here, followed by an irreversible chemical reaction,
as detailed in the next paragraph.
At non-zero probe concentrations, we describe physical probe-RNA interactions through a thermodynamic binding model with  two trainable parameters ($\mu_r$ and $\Delta\mu_{\text{pairing}}$).
The reference chemical potential $\mu_r$ captures the probability of an unpaired nucleotide to bind to a chemical probe
at a reference DMS concentration $[\textrm{DMS}]_r=1\textrm{M}$.
The chemical potential is then corrected based on the actual probe concentration [DMS] as $\mu = \mu_r + \mathrm{k_BT}\ln\left(\frac{[\text{DMS}]}{[\text{DMS}]_r}\right)$.
$\Delta\mu_{\text{pairing}}$ differentiates between paired and unpaired sites, so that
the effective chemical potential for paired nucleotides becomes $\mu^\prime\equiv\mu - \Delta\mu_{\text{pairing}}$.
Since chemical probes typically exhibit reduced accessibility to base-paired regions, $\Delta\mu_{\text{pairing}}$ is expected to be positive,
though the model could be straightforwardly generalized to cases where this is not true.
Incorporating a penalty term for base-paired sites results in a modified partition function.
This modification is implemented by applying a concentration-dependent energetic contribution to paired bases
$\Delta F_{\text{[DMS]}}(s)$
(derivation is detailed in the SI Appendix).

Putting together all these terms, the corrected free energy of a given secondary structure $s$ is
\begin{align}
    F(s) &= \underbrace{F_0(s)}_{\text{baseline}} + \underbrace{\sum_{i \in \text{paired}(s)} \lambda_i}_{\substack{\text{site-specific}\\ \text{corrections}}} + \underbrace{\Delta F_{\text{[DMS]}}(s)}_{\substack{\text{concentration} \\ \text{dependent perturbation}}}
        \label{eq:free_energy_maintext}
\end{align}
from which we can compute the population of observing each structure $s$ via \autoref{eq:ps_maintext}.
\paragraph*{\textit{Chemical binding}}
Following the physical interaction between a chemical probe and RNA, adducts are formed with probabilities that depend on both nucleobase identity and local RNA structure, under fixed experimental conditions (e.g., temperature and probing time).
In the most general framework, this yields eight possible configurations, defined by the combination of nucleotide type (A, U, C, or G) and structural state (paired or unpaired).
For DMS probing, due to the position of the potentially reactive nitrogens,
adenine and cytosine show negligible reactivity when base-paired, while guanine and uracil retain reduced but secondary structure-independent reactivity.
Accordingly, the model is parameterized by four effective binding probabilities, corresponding to the reactive nucleotide-structure combinations:
$p_{\text{bind}}(\text{A},\text{unpaired})$, $p_{\text{bind}}(\text{C},\text{unpaired})$, $p_{\text{bind}}(\text{G})$, and $p_{\text{bind}}(\text{U})$.

\subsubsection*{Mutational Profiling and prediction of reactivity profiles}\label{subsubsec:mutational_profiling}
\paragraph*{\textit{Mutational profiling}}
Following chemical probing, modifications are typically detected through reverse transcription.
Here, we focus on mutational profiling (MaP), though only minor modifications to the model would be required for compatibility with other experimental techniques, such as RT-stop.
During reverse transcription, we account for the probability of producing false positive or false negative signals through the parameters $m_0$ and $m_1$, as detailed in the SI Appendix.
Applying this framework to a given secondary structure $s$, we can compute the expected reactivity profile for that structure, then sum over all secondary structures in the ensemble to obtain the expected reactivity profile for the experiment.
\paragraph*{\textit{Handling systematic errors}}
Sequencing errors are known to introduce systematic biases in reactivity profiles \citep{liu-weiSequencingAccuracySystematic2024}, and reverse transcription can be affected by the presence of secondary structures \citep{verwiltArtifactsBiasesReverse2023}.
To account for such effects, we make use of experimental controls (performed without reagent) to estimate position-dependent correction factors $\epsilon_i$, which effectively act as a baseline adjustment to the predicted mutation rates at each nucleotide position and are computed so that the prediction at zero concentration of DMS matches the control data.
These factors capture systematic biases that are sequence-specific but independent of probe concentration.
This allows the model to focus on the concentration-dependent signal in reactivity, which is most informative about RNA structure and probe interactions, while $\epsilon_i$ absorbs background effects that remain constant across conditions.
In other words, the sequence-dependent $\epsilon_i$ parameters model the mutation rate in absence of reagents, whereas the
sequence-dependent soft constraints $\lambda_i$ affect the pairing population which, in turn, controls how much the mutation rate increases
when the reagent is added.
As discussed in the SI Appendix, when control experiments are available the parameter $m_0$ is omitted, since background correction factors can be directly estimated from the control data. If controls are unavailable, $m_0$ is instead fitted to provide a single background correction factor shared across the sequence.

\subsection*{Parameter Optimization}\label{subsec:fit_parameters}

Upon computation of the expected reactivity profile for a given sequence and probe concentration, we optimize the model parameters by minimizing a loss function that quantifies the discrepancy between predicted and observed reactivities.
We model per-position mutation counts as binomial: given read depth $n_i$ and model-predicted mutation probability $M_i$, each read is an independent Bernoulli trial. Accordingly,
\begin{align}
    \mathbb{P}\big(\mathcal{M}_i^{\mathrm{exp}} \mid n_i, M_i\big)
    &= \mathrm{Binomial}\big(\mathcal{M}_i^{\mathrm{exp}}; n_i, M_i\big)
\end{align}
where $\mathcal{M}_i^{\mathrm{exp}}$ is the experimentally observed number of mutations at position $i$.
As discussed in the SI Appendix, the overall likelihood of the model across all nucleotide positions is the product of the individual position likelihoods.
To facilitate optimization, we minimize the negative log-likelihood, yielding the following loss function:
\begin{align}
    \mathcal{L}
    &= - \sum_i \ln \, \mathbb{P}\big(\mathcal{M}_i^{\mathrm{exp}} \mid n_i, M_i\big).
    \label{eq:loss_maintext}
\end{align}
The derivation and all definitions are expanded in the SI Appendix.

This optimization is implemented via a gradient-based algorithm that systematically adjusts model parameters to minimize the discrepancy between predicted and experimental reactivity profiles \cite{byrdLimitedMemoryAlgorithm1995} as implemented in SciPy
\cite{virtanenSciPy10Fundamental2020}.
The parameter optimization is performed in two consecutive phases, as illustrated in \autoref{fig:1}.
In the first phase, we optimize the global physical parameters that are shared across all sequences and concentrations.
These include: the reference chemical potential ($\mu_r$), the binding penalty for structured regions ($\Delta\mu_{\text{pairing}}$), the nucleotide-specific adduct formation probabilities ($p_{\text{bind}} (\text{A},\text{unpaired})$, $p_{\text{bind}} (\text{C},\text{unpaired})$, $p_{\text{bind}} (\text{G})$, and $p_{\text{bind}} (\text{U})$), the false negative rate parameter ($m_1$), and—when experimental controls are unavailable—the background correction factor ($m_0$).
This phase optimizes a total of 7 parameters (8 if no controls are available).
In the second phase, we look to refine the secondary structure ensemble to find the one that best explains the experimental data.
This is achieved by optimizing the sequence-specific soft constraints ($\lambda_i$) for each nucleotide position, while keeping the global physical parameters fixed.

Following parameter optimization, $\Delta F_{\text{[DMS]}}(s)$ is set to zero,  extrapolating to zero probe concentration and thereby revealing the ensemble of secondary structures most consistent with the experimental data under native conditions.

\subsection*{Experimental procedures}

DMS probing of a synthetic RNA construct was performed as described in the appendix.
DMS-treated RNA was column-purified and subjected to reverse transcription using MarathonRT \cite{zhaoUltraprocessiveAccurateReverse2018,guoSequencingStructureProbing2020,bohnNanoDMSMaPAllowsIsoformspecific2023,gribling-burrerIsoformspecificRNAStructure2024}.
Sequencing reads were processed with an RNAFramework-based workflow \cite{incarnatoRNAFrameworkAllinone2018,langmeadFastGappedreadAlignment2012,danecekTwelveYearsSAMtools2021}.

\section*{Results}\label{sec:results}

\subsection*{Transferability and accuracy of the physical model}\label{sec:res1}
We extracted physical parameters from our unified physics-based framework after training on a set of well-characterized RNA structures previously studied using DMS chemical probing \citep{bohnNanoDMSMaPAllowsIsoformspecific2023}.
This set spans a range of structural complexity, comprising five distinct systems: (i) bacterial RNase P (type A), (ii) hc16 ligase, (iii) \textit{Tetrahymena} ribozyme, (iv) \textit{V. cholerae} glycine riboswitch, and (v) HCV IRES. These sequences range from approximately 100 to 400 nucleotides in length.
Because all systems were probed under consistent conditions, with multiple DMS concentrations and independent replicates, this dataset is well suited for testing whether the physical parameters of the model transfer across sequences rather than being overfitted to any single system.
Reference structures for hc16 ligase, \textit{Tetrahymena} ribozyme, and \textit{V. cholerae} glycine riboswitch were obtained from cryo-EM-guided structure determination \citep{kappelAcceleratedCryoEMguidedDetermination2020}, and secondary structure was annotated from the 3D structure with Barnaba (v0.1.8) \citep{bottaroBarnabaSoftwareAnalysis2019}. HCV IRES reference structure was manually transcribed from Extended Data Fig. 6 of Ref.~ \citep{bohnNanoDMSMaPAllowsIsoformspecific2023}. For bacterial RNase P (type A), although a crystal structure exists \citep{torres-lariosCrystalStructureRNA2005}, sequence differences between the crystallized construct and the chemically probed sequence preclude its use as a reference.

These references are expected to be more accurate than predictions from a simple thermodynamic model, but since they represent single conformations, they cannot by themselves report on conformational heterogeneity.

\begin{figure}[!tbp]
\centering
\includegraphics[width=\linewidth]{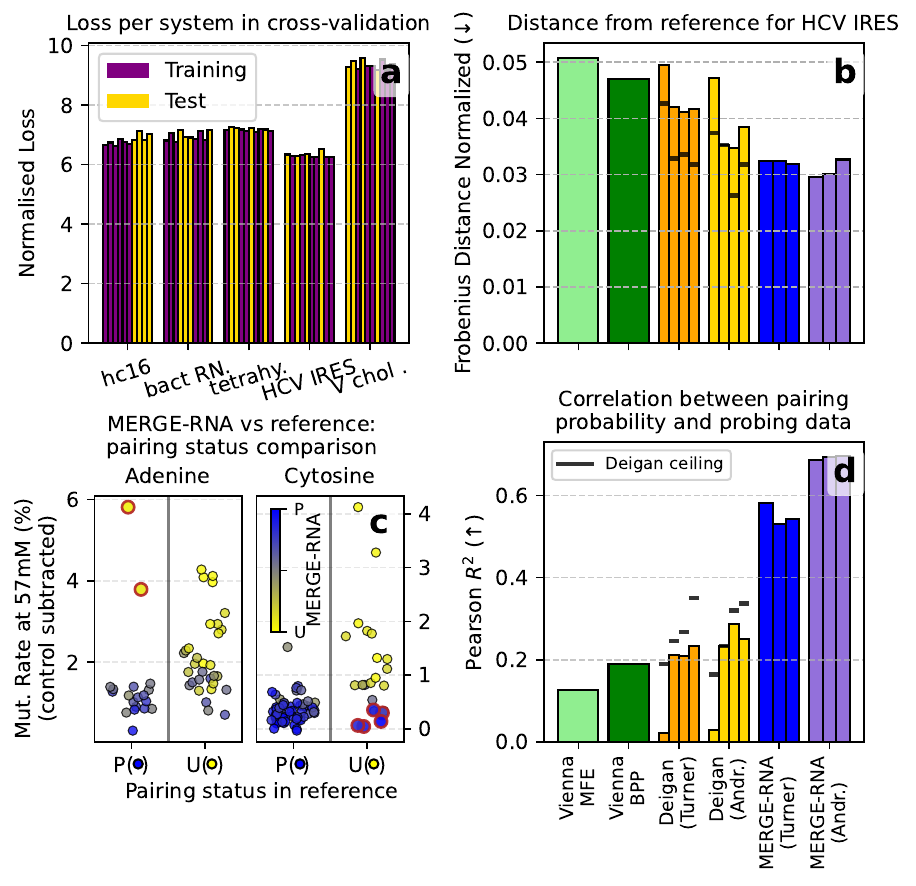}
\caption{
\textbf{(a)} Cross-validation of physical parameters across five RNA systems.
Each bar represents the per-datapoint loss obtained for the indicated system after training on three systems and testing on the remaining two.
Yellow bars refer to results obtained when the corresponding system was excluded from the training set.
\textbf{(b)} Frobenius distance divided by the squared number of nucleotides between base pairing probability (bpp) matrices, quantifying structural similarity between the HCV IRES reference structure and: MFE structure, bpp of the original thermodynamic ensemble,
bpp obtained from standard pseudo-free-energy method (Deigan) and bpp from MERGE-RNA.
Deigan results are reported using different DMS concentrations, and
MERGE-RNA results are reported for different replicates (independent fits).
Deigan and MERGE-RNA results are reported for different baseline thermodynamic models (Turner and Andronescu).
Solid bars for Deigan method use $(m,b)$ coefficients determined with a standard jackknife procedure, while the caps mark its
theoretical best performance (see text for definition).
Colors as in panel (d).
\textbf{(c)} Mutation rates at 57mM DMS concentration after background subtraction for adenine and cytosine in HCV IRES, grouped by pairing status in the reference structure. Each point is colored according to the pairing probability predicted by MERGE-RNA. Strong disagreements (positions where MERGE-RNA predicts pairing probability below 25\% or above 75\% while the reference indicates the opposite) are highlighted.
\textbf{(d)} Squared Pearson correlation coefficient between predicted base pairing probabilities and observed mutation rates, computed separately for A and C and averaged.
The caps show the theoretical best performance of the Deigan method.
Labels as in panel (b).
}
\label{fig:redmond}
\end{figure}
\begin{figure}[!tbp]
\centering
\includegraphics[width=\linewidth]{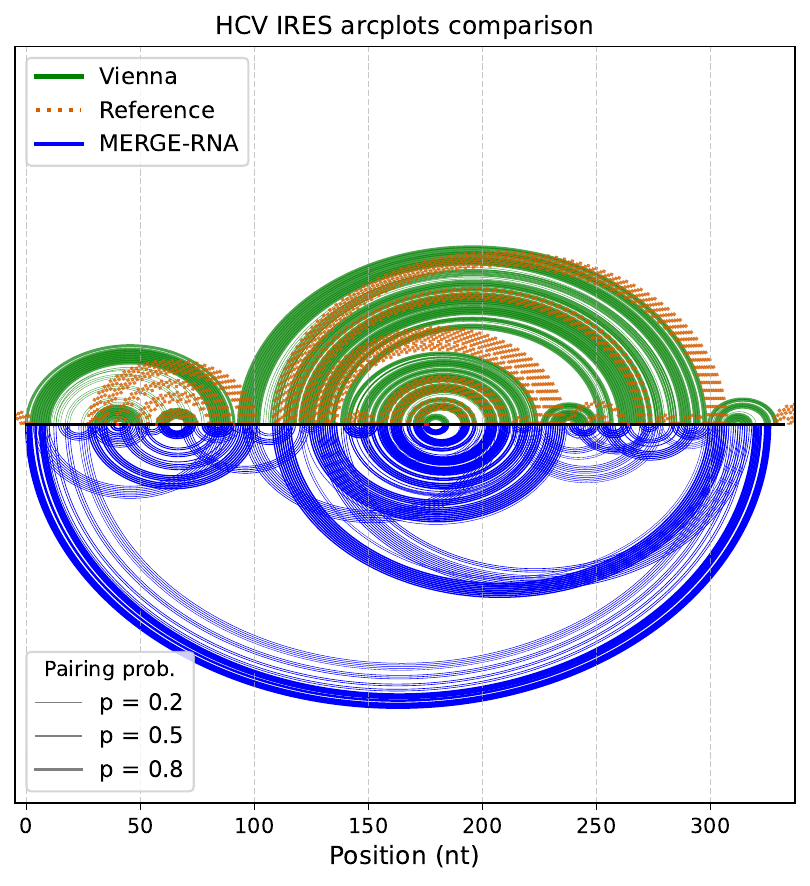}
\caption{Arc plot visualization comparing the secondary structures of HCV IRES: reference experimental structure (red), ensemble obtained from the thermodynamic model (green), and ensemble obtained from MERGE-RNA (blue).
}
\label{fig:arcplot}
\end{figure}

Our first objective was to assess whether the physical parameters extracted from our model represent genuine physical properties of the experiment and are hence transferable to different systems, rather than being overfitted on a specific case.
We performed comprehensive cross-validation fits, training only the physical parameters of the model (all the parameters except the $\lambda_i$, which are fixed to 0 for the moment) on triplets of RNA systems and testing on the remaining pairs (Fig.~\ref{fig:redmond}a).
Loss values (as defined in \siautoref{eq:loss}) are then divided by the number of data points to enable direct comparison.
Obtained values of the physical parameters are reported in \siautoref{tab:physical_params} for each fit.
The loss values obtained across all sets are consistent both in training and in test, demonstrating that our model generalizes on multiple systems and successfully captures the underlying principles governing chemical probing experiments with a small set of parameters.

\subsection*{Inference of optimal ensemble}

Before turning to the full inference of the per-nucleotide soft constraints, we verified with a simple controlled experiment, reported in the SI Appendix ("\nameref{si:heterogeneity}", Supplementary Figs. \siref{fig:fig2b_HCV_IRES}--\siref{fig:fig2b_hc16}), that a proper balance between reference structural information and ensemble heterogeneity is needed to reproduce chemical-probing data: applying a uniform reference-derived soft constraint of increasing magnitude reveals a ``sweet spot'' where the model partially incorporates the reference while preserving the conformational diversity captured by the experiment, beyond which performance degrades.
Building on this insight, instead of borrowing from the reference structure, we inferred the best sequence-specific soft constraints ($\lambda_i$ parameters) using our model. This resulted in substantially improved agreement with experimental data (blue line in Figs. \siref{fig:fig2b_HCV_IRES}--\siref{fig:fig2b_hc16}), as expected given that the experimental data are used to infer the soft constraints.

To validate our model independently, we computed the Frobenius distance between the reference structure and base-pairing probability matrices obtained by the baseline ViennaRNA MFE, ViennaRNA BPP, and by standard pseudo-free energy approach as in Ref. \cite{deiganAccurateSHAPEdirectedRNA2009} (Fig.~\ref{fig:redmond}b and Figs.~\ref{fig:fig2_tetrahymena_ribozyme}--\ref{fig:fig2_hc16}).
For the Deigan approach, rather than using the literature coefficients $(m,b)=(1.8,-0.6)$, for each DMS concentration we selected the best-fitting slope and intercept through a procedure analogous to \texttt{rf-jackknife} \citep{morandiGenomescaleDeconvolutionRNA2021, incarnatoRNAFrameworkAllinone2018}, using a leave-one-out approach to avoid fitting these coefficients on the same structural data we are optimising, as detailed in the SI Appendix (\nameref{si:deigan_refit}).
Solid bars report results using these coefficients, while the dashed caps mark the best value the Deigan functional form can reach (its ceiling), i.e. the value obtained by choosing $(m,b)$ to maximize the agreement with the metric itself.
To test robustness with respect to the choice of thermodynamic baseline, we repeated the analysis using both the standard Turner2004 \cite{mathewsIncorporatingChemicalModification2004}, and the Andronescu2007 parameters \cite{andronescuEfficientParameterEstimation2007}.
For ensemble-based metrics, our model improves significantly over Deigan's in three of the four systems studied, and in general remains comparable to the theoretical best results achievable by Deigan's method, obtained when it is fitted to the reported test data
(Figs. \siref{fig:fig2_tetrahymena_ribozyme}--\siref{fig:fig2_hc16}, in SI Appendix).
This is remarkable, considering that MERGE-RNA is not optimized for structure prediction at any stage of its pipeline.
We also report in the SI Appendix (Figs. \siref{fig:deigan_metrics_HCV_IRES}--\siref{fig:deigan_metrics_V_chol_gly_riboswitch}) metrics more commonly reported for single-structure prediction: positive predictive value, sensitivity, and the same Frobenius distance computed on the MFE structure rather than on the ensemble in order to isolate the effect of selecting only the MFE from the ensemble.
When evaluating the MFE structure, the Deigan approach is favoured.
This is expected, since the Deigan pseudo-free-energy approach is explicitly designed to improve single-structure prediction, whereas MERGE-RNA is optimized to infer an ensemble consistent with the probing data rather than to maximize agreement of any single structure with the reference.
Arc plots (Fig.~\ref{fig:arcplot}) illustrate how MERGE-RNA's individualized fitting preserves ensemble diversity (blue) while recapitulating the reference structure (red) more accurately than the standard thermodynamic ensemble (green).
The same comparison with the Deigan ensemble superimposed in place of the unconstrained thermodynamic one, for each of the four systems, is reported in the SI Appendix (Figs. \siref{fig:full_seq_arcplot_HCV_IRES_deigan}--\siref{fig:full_seq_arcplot_V_chol_gly_riboswitch_deigan}).
Similar improvements were observed across all tested RNA systems, confirming the broad applicability of our approach.

\subsection*{MERGE-RNA suggests localized structural dynamics supported by chemical probing data}
Having established structural accuracy by comparison with experimentally resolved reference structures, we now examine where our predictions diverge from the static reference.
Since our model predicts a dynamic ensemble rather than a single conformation, divergence is expected and informative: chemical probing experiments inherently report on the dynamic ensemble of the RNA molecule, which a static reference structure alone cannot fully capture.
In \autoref{fig:redmond}c we report mutation rates after control subtraction for adenine and cytosine, grouped by pairing status in the reference structure.
As expected, paired positions show, on average, lower mutation rates than unpaired ones.
Each point is colored according to the pairing status predicted by MERGE-RNA.
The model's predictions correlate well with mutation rates, demonstrating its capability to infer and model pairing dynamics from probing data.

It is instructive to examine positions where our predictions disagree with the reference structure.
When mutation rates are consistent with the reference structure, MERGE-RNA's predictions align with it.
For positions with intermediate mutation rates, MERGE-RNA predicts intermediate pairing probabilities, suggesting conformational heterogeneity in solution.
This is particularly informative: a static reference structure classifies each position as either paired or unpaired, and thus cannot account for intermediate reactivity.
MERGE-RNA provides an alternative explanation for this discrepancy by proposing alternative conformations that together explain the observed mutation rates.
Where probing data and reference structure are in contrast, MERGE-RNA captures new pairings absent from the reference but supported by probing, or identifies reference-annotated pairs that show elevated mutation rates, suggesting they are frequently unpaired in the dynamic ensemble.

The number of ``strong disagreements sites'' (positions where MERGE-RNA predicts pairing probability below 25\% or above 75\% while the reference indicates the opposite for A and C nucleotides, highlighted in Fig.~\ref{fig:redmond}c) is modest, ranging from 0 to 13 across the systems studied here.
Detailed analysis (SI Appendix Figs. \siref{fig:full_seq_arcplot_HCV_IRES_detailed}--\siref{fig:full_seq_arcplot_V_chol_gly_riboswitch_detailed}) reveals that most correspond to small corrections of existing structural elements or formation/dismantling of short motifs (e.g., 3--4 bp stems), indicating localized inconsistencies rather than wide structural disagreements.
Some level of disagreement is expected, as the reference is a single structure obtained under specific conditions (e.g., crystallographic, or with a different buffer), whereas probing reflects the ensemble populated under the experimental conditions of the probing itself, which in principle can be very different; MERGE-RNA adjusts the reference only where the probing data require, a behavior that becomes valuable when the reference is partial or originates from conditions that differ from those of the probing experiment.
The correspondence between base-pairing probability and observed mutation rates can be quantified by computing
their Pearson correlation (Fig.~\ref{fig:redmond}d): MERGE-RNA shows a noticeably higher correlation than the one obtained by the traditional pseudo-free energy approach across all systems.
This gap is not a matter of how the $(m,b)$ Deigan coefficients are chosen.
In fact, MERGE-RNA exceeds the results obtained with the coefficients cherry-picked to maximize the displayed metric
by $\Delta R^2 = +0.18$ in the worst case and $+0.28$ on average (1.4x and 1.9x, respectively) across all systems and concentrations.
This indicates that the gap stems from the limited functional form and degrees of freedom of the two-parameter pseudo-energy scheme, rather than from a suboptimal choice of coefficients or the data.

In summary, with our model
(1) the inferred physical parameters are generalisable across systems (Fig.~\ref{fig:redmond}a);
(2) we show that there is an optimal balance between structural accuracy and ensemble heterogeneity, and inference of sequence-specific corrections further improves agreement with experimental data (Fig.~\siref{fig:fig2b_HCV_IRES}--\siref{fig:fig2b_hc16});
(3) the resulting structural ensembles more accurately capture native conformations compared to standard thermodynamic approaches (Fig.~\ref{fig:redmond}b);
(4) the resulting ensemble correlates significantly better with the observed mutation rates than traditional methods (Fig.~\ref{fig:redmond}d);
(5) comparing MERGE-RNA's results to the reference structures reveals that agreement reflects probing data consistency (Fig.~\ref{fig:redmond}c).
In cases where reference models are inconsistent with experiment, MERGE-RNA goes beyond the static reference by identifying localized discrepancies and proposing alternative conformations that better explain the data.
These results are consistent across all tested RNA systems (SI Appendix Figs. \siref{fig:fig2_tetrahymena_ribozyme}--\siref{fig:fig2_hc16}).
However, in the absence of independent experimental evidence for ensemble populations, we cannot quantitatively validate the inferred alternative conformations here; in particular, we cannot rule out that MERGE-RNA over-estimates ensemble heterogeneity by accommodating intermediate reactivities with intermediate pairing probabilities. For this reason, we next turn to the adenine riboswitch, where independent NMR measurements provide ground truth on both the co-existing conformers and their ligand-dependent populations.

\subsection*{Application on a known case of functional structural rearrangement: \textit{V. vulnificus add} adenine riboswitch}\label{sec:adenine_riboswitch}

\begin{figure}[!tbp]
\centering
\includegraphics[width=\linewidth]{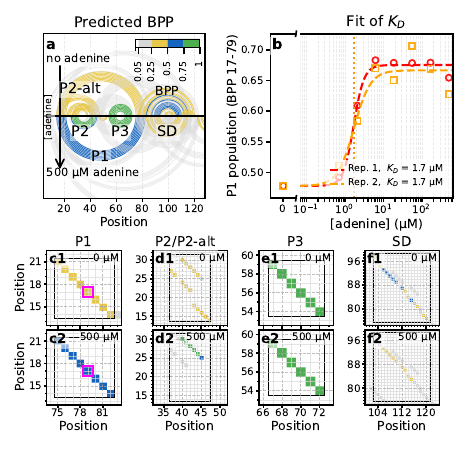}
\caption{
MERGE-RNA captures ligand-induced structural rearrangements in the \textit{V. vulnificus add} adenine riboswitch.
\textbf{(a)} Arc plot of the inferred base-pairing probabilities at 0 µM (top) and 500 µM (bottom) adenine; colors indicate base-pairing probability.
\textbf{(b)} Estimation of the ligand-binding dissociation constant ($K_d$) from the inferred pairing probability of a representative base pair in the P1 helix (highlighted with a purple box in c1--c2), which is stabilized upon ligand binding.
\textbf{(c1--2, d1--2, e1--2, f1--2)} Zoomed views of the predicted BPP matrices for the P1, P2, P3, and SD domains in the absence (top row) and presence (bottom row) of adenine. Color scale as in panel (a).
}
\label{fig:adenine_riboswitch_maintext}
\end{figure}

Having established structural accuracy and parameter transferability on structured RNAs, we now turn to the validation of the model's ability to deconvolve co-existing conformational states.
For this purpose we use the \textit{Vibrio vulnificus add} adenine riboswitch, which undergoes a well-characterized ligand-induced conformational change that regulates translation, and provides independent experimental ground truth unavailable for the structured RNAs above: its ligand-free and ligand-bound conformations and the binding dissociation constant have been resolved by NMR \cite{reiningThreestateMechanismCouples2013}, and the system has served as a benchmark for other ensemble-deconvolution methods including DRACO \cite{morandiGenomescaleDeconvolutionRNA2021} and DANCE-MaP \cite{olsonDiscoveryLargescaleCellstateresponsive2022}.

We analyzed previously deposited DMS-MaP data acquired across a range of adenine concentrations, from 0 to 500 µM \cite{olsonDiscoveryLargescaleCellstateresponsive2022}.
Since the experiment provides only a single DMS concentration (170mM) alongside the no-DMS control, we resorted to the linear version of the model, which assumes that the probe perturbs pairing probabilities negligibly and that the probe concentration is well below saturation, which corresponds to the regime of low chemical potential of probe-RNA interactions, avoiding fitting a nonlinear model to only two points.
Because our model does not explicitly account for the reduction in mutation rate of unpaired bases caused by ligand contacts and does not model pseudoknots, we masked respectively the ligand-binding pocket (J) and the loops (L2 and L3), to prevent the model from fitting these trends.
For the same reason, the physical parameters of the model were fit using only the ligand-free experiments; we then optimized the $\lambda_i$ parameters separately for each condition to infer the structural ensemble at every adenine concentration.

We present in Figure~\ref{fig:adenine_riboswitch_maintext} the inferred ensembles at the two limiting conditions, ligand-free (0 µM adenine) and ligand-saturated (500 µM adenine). The full concentration series is provided in Figure~\siref{fig:adenine_riboswitch_arcgrid} and smoothly interpolates between these two points. Panel~(a) shows the arc-plot of the base-pairing probabilities, and panels (c1--2, d1--2, e1--2, f1--2) provide zoomed views of the predicted BPP matrices for the P1, P2, P3, and SD domains in the absence (top row) and presence (bottom row) of adenine.

The model recovers the molecule's structural details and the expected ligand-induced rearrangement: the P1 helix is stabilized upon ligand binding; the P2 region transitions from a mixture of alternative pairings (P2/P2-alt) to a single stabilized P2 conformation; the P3 helix remains stable in both conditions; and the SD domain (which harbors the Shine-Dalgarno sequence) is destabilized upon ligand binding, consistent with the known mechanism of translational regulation by this riboswitch.
To validate against the independent NMR data, similarly to what has previously been done by DANCE-MaP authors, we estimated the dissociation constant from the ligand-dependent pairing probability predicted by MERGE-RNA of a representative base pair in the P1 helix (17--79, highlighted with a purple box in c1--c2) that is stabilised when ligand is present.
We tracked this pairing probability across all adenine concentrations and fit it to a Hill curve,
\begin{equation*}
y(x) = y_{\min} + (y_{\max} - y_{\min})\,\frac{x^n}{K^n + x^n},
\end{equation*}
where $x$ is the adenine concentration (in µM), the $0$ µM point is pinned as the lower plateau, and $K$ and $n$ are estimated by bounded nonlinear least squares.
The midpoint concentration $K$ is $K_d = 1.7$ µM for both the independent fits, in good agreement with the NMR-derived value of 1.6 µM \cite{reiningThreestateMechanismCouples2013} and the previous prediction of $1.5 \pm 0.3$ µM reported in ref.~\cite{olsonDiscoveryLargescaleCellstateresponsive2022}.

Taken together, MERGE-RNA reproduces essentially all features of the adenine riboswitch that are accessible in NMR experiments \cite{reiningThreestateMechanismCouples2013}: the three resolved structures, their ligand-dependent rearrangement, and the dissociation constant $K_d$. The only structural element that is, by construction, out of reach for our model is the pseudoknot involving L2 and L3, since the underlying thermodynamic baseline does not allow pseudoknotted pairings.
For comparison, we repeated the same analysis using the Deigan pseudo-free-energy approach (Figs.~\ref{fig:adenine_riboswitch_maintext_deigan}--\ref{fig:adenine_riboswitch_arcgrid_deigan}). While the optimized Deigan model predicts a similar apparent $K_d$, it substantially over-stabilizes the P1 helix (predicted population always $>97\%$) and fails to reproduce the ligand-dependent destabilization of the Shine–Dalgarno sequence.

Beyond the conformations resolved by NMR, MERGE-RNA also predicts intermediate states along the P1$\leftrightarrow$SD competition (P4 in Ref.~\cite{reiningThreestateMechanismCouples2013}), in the form of a gradient of pairing probabilities within the two competing helices rather than only the fully-formed alternatives (Fig.~\siref{fig:adenine_riboswitch_arcgrid}).
We cannot directly validate this prediction against the NMR data, for two reasons.
First, the construct used in the probing experiment is longer than the one resolved by NMR: it carries additional upstream nucleotides that can pair with the 3$^\prime$ side of P1, stabilizing a longer version of the helix (``P1-extended'' in Ref.~\cite{olsonDiscoveryLargescaleCellstateresponsive2022}).
Second, NMR structure determination tends to discretize the conformational space into a small number of representative structures, so its absence in the NMR ensemble cannot be taken as evidence against the existence of intermediate populations.
That said, the existence of intermediates is consistent with the NMR conformers themselves, which resolve P1 with 0, 4, and 7 base pairs in apo-B, apo-A, and holo, respectively, three discrete points along what is plausibly a continuous distribution of partially formed P1 helices.
Read-clustering methods such as DRACO \cite{morandiGenomescaleDeconvolutionRNA2021} and DANCE-MaP \cite{olsonDiscoveryLargescaleCellstateresponsive2022} similarly discretize the ensemble into a small number of representative structures and therefore struggle to describe such a continuum, whereas MERGE-RNA naturally captures it.

We also note a methodological distinction: DRACO and DANCE-MaP exploit the co-occurrence of modifications within individual reads to assign each read to a discrete cluster, which requires deep coverage and effectively discards reads with zero or a single modification.
MERGE-RNA instead operates on the cumulative mutation profile alone, so every read contributes to the inference and the per-position pairing probabilities are recovered without any per-read correlation analysis.
The population gradient observed in the P1$\leftrightarrow$SD competition is reminiscent
of a strand-displacement. Interestingly, strand-displacement has been suggested as a possible mechanism to control the expression-platform conformation in transcriptional riboswitches \cite{chengCotranscriptionalRNAStrand2022}. Strand-displacement is also observed in a more controlled setting using new DMS data below (see Section ``A putatively bistable sequence exhibits a heterogeneous ensemble with strand displacement'').
We also note that although the data were collected with four-base DMS \cite{mustoeRNABasepairingComplexity2019}, our framework currently treats only A and C as pairing-sensitive reactive bases.
Although this reduces the information used relative to the original four-base DMS analysis, the A/C-sensitive model is sufficient to recover the main ligand-dependent rearrangement.
Extending the model to all four bases is a natural direction for future work.

Overall, this case study shows that, from chemical probing data alone, MERGE-RNA resolves the structural detail of multiple competing conformers identified by NMR in a biologically relevant RNA. Additionally, it captures the ligand-induced shift in the conformational ensemble, quantitatively matching the NMR-derived dissociation constant.
Importantly, conformational ensembles of this riboswitch were here controlled by ligand concentration alone, which allowed a series of experiments performed in similar conditions. The more challenging case of a thermoswitch, which requires combining experiments at different temperatures and acquisition times \citep{giuliodoriCspAMRNAThermosensor2010,zhangStressResponseThat2018}, is reported in Supplementary Information.

\subsection*{Deconvolution of mixed structural states on synthetic data}\label{sec:synthetic_data}

\begin{figure}[!tbp]
\centering
\includegraphics[width=\linewidth]{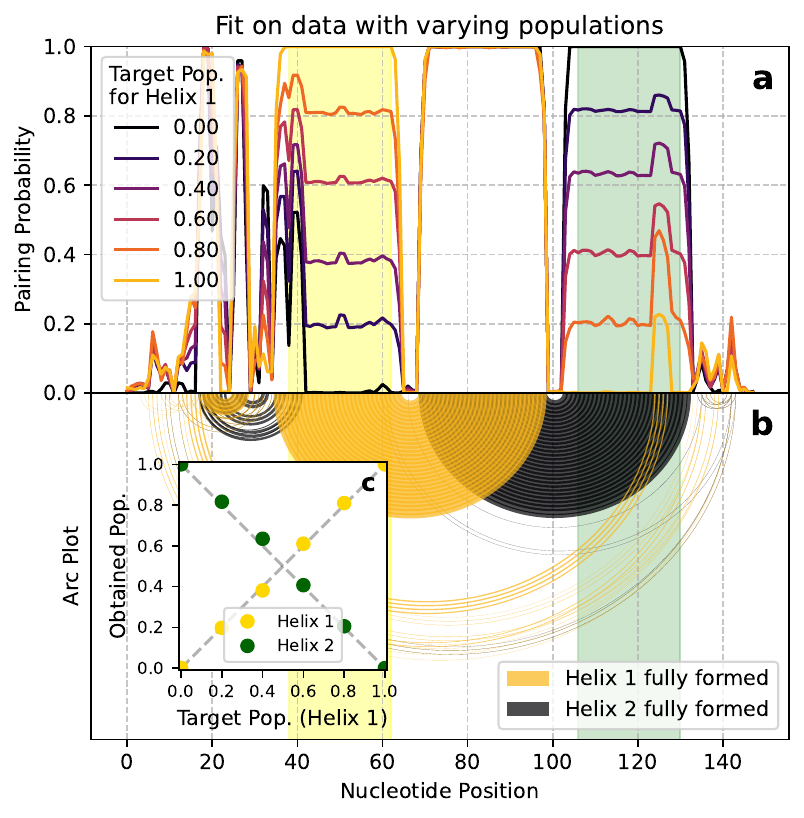}
\caption{
Model accurately deconvolves mixed structural states from synthetic data.
\textbf{(a)} Base-pairing probabilities inferred from fitting on synthetic data for a bistable RNA sequence.
The data were generated from an RNA construct with two competing helices (helix 1 and helix 2), where the ground-truth population of helix 1 was systematically varied from $0$ to $1$ with steps of $0.2$.
The model's predicted pairing probabilities are shown for each case.
\textbf{(b)} Arc plots of the inferred structural ensemble for two limit cases: when helix 1 is dominant (gold) and when helix 2 is dominant (black).
\textbf{(c)} The helix population inferred by the model is plotted against the ground-truth population.
The inferred population is quantified as the median of the base-pairing probabilities over the respective helical regions, highlighted in panels (a) and (b).
}
\label{fig:fig3}
\end{figure}

To rigorously test the model's ability to quantitatively predict secondary-structure populations in heterogeneous ensembles, we generated synthetic datasets for an in-house designed bistable RNA carrying two mutually exclusive hairpins.
We deliberately designed this construct so that we could also assay the very same sequence experimentally (discussed in the next section); accordingly, we appended primer binding sites at the 3$^\prime$ and 5$^\prime$ ends.
Concretely, the final construct has the architecture
\begin{align}
	5^\prime\text{--PBS1--}\text{A1--L1--B}\text{--L2--A2}\text{--PBS2--}3^\prime
		\label{eq:seq_design}
\end{align}
where PBS1 and PBS2 are distinct primer-binding sites, A1 and A2 are identical copies of the same sequence, and B is complementary to A1/A2. In this design, A1 and A2 compete to pair with B, producing two mutually exclusive hairpins (helix 1: A1:B; helix 2: A2:B). L1 and L2 are short linkers designed to act as tetraloops for the hairpins.

While the original design targeted roughly equal populations of the two helices, the thermodynamic model allows us to artificially bias the ensemble to favor either helix.
We then simulated mutation profiles for the two cases—when either helix is always formed—and combined these profiles in known proportions to create synthetic datasets with varying helix populations, ranging from $0$ to $1$ in increments of $0.2$.
To generate realistic mutation profiles, we used the physical parameters inferred from our analysis on experimental data from biologically relevant RNAs (subsection 'Transferability and accuracy of the physical model' above).

We then trained our model on these synthetic datasets.
Following the procedure established earlier, we first performed a joint fit of the physical parameters across all datasets, starting from random initializations.
Subsequently, we optimized the soft constraints $\lambda_i$ for each dataset individually.
The results, presented in Fig.~\ref{fig:fig3}, show that the model accurately recovers the expected base-pairing probabilities for both helices across all synthetic datasets. The method is robust to the presence of additional base pairs formed by the random primer binding sites.
Overall, these synthetic data tests validate the model's ability to recover structural populations consistent with the known ground truth (Fig.~\ref{fig:fig3}c).

\subsection*{A putatively bistable sequence exhibits a heterogeneous ensemble with strand displacement}\label{sec:newseq}

\begin{figure}[!tbp]
\centering
\includegraphics[width=\linewidth]{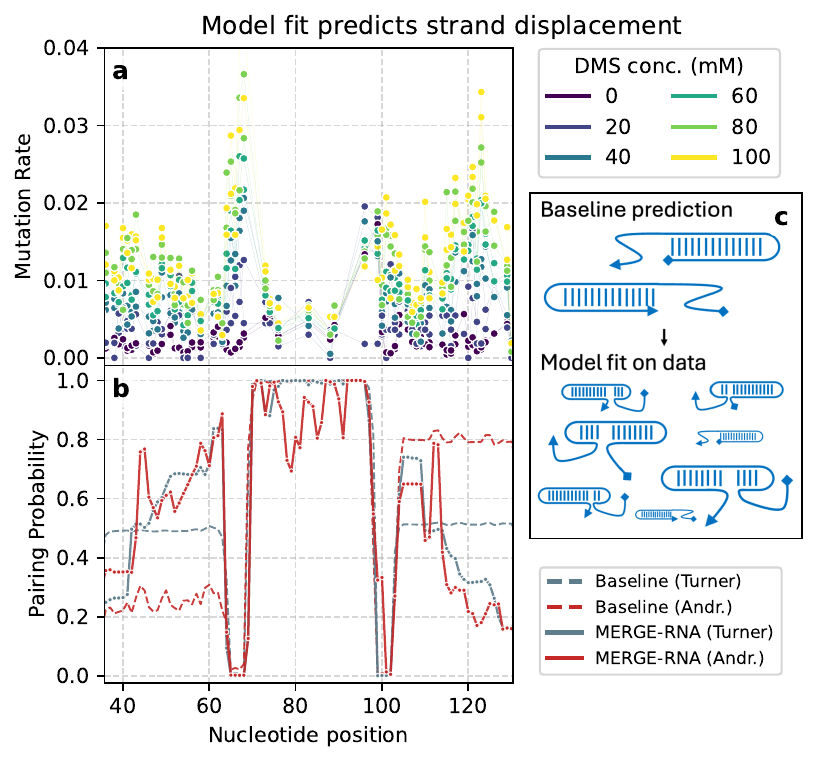}
\caption{Ensemble inference on experimental data for a putative bistable RNA exhibits evidence of strand displacement.
	\textbf{(a)} Per-nucleotide DMS-MaP mutation profiles for adenines and cytosines across 0--100 mM (20 mM steps; two replicates each) for a putative bistable RNA. We leave out of the plot the primer binding sites for better visualization. Two regions (A1 on the left and A2 on the right, as defined in (\ref{eq:seq_design})) are competing to pair with the central region (B). They show a slope in the reactivity profile that is compatible with a strand displacement mechanism. \textbf{(b)} Pairing probabilities obtained from the ensemble predicted by MERGE-RNA (solid lines) after fitting on experimental data, with Turner 2004 \citep{mathewsIncorporatingChemicalModification2004} (crimson) or Andronescu 2007 \citep{andronescuEfficientParameterEstimation2007} (steel) thermodynamic models as baselines. MERGE-RNA is able to fix both the baseline predictions (dotted lines) and retrieve pairing probabilities consistent with the observed mutation rates. \textbf{(c)} Schematic depiction of the ensemble of structures as predicted by the baseline thermodynamic model (above) and MERGE-RNA fitted on the experimental data (below). While the thermodynamic model predicts only the two configurations with fully-formed helices, MERGE-RNA captures a wealth of intermediate states where strand displacement occurs.}
\label{fig:fig_newseq}
\end{figure}

To validate our model on experimental data for the same system, we collected DMS-MaP data on the putative bistable RNA construct described in the previous section.
The data were collected at multiple DMS concentrations, 0 to 100 mM at steps of 20 mM, 2 replicates each.
The obtained mutation profiles are visualized in Fig.~\ref{fig:fig_newseq}a.
We then trained our model on these experimental datasets, following the same procedure as in the previous section.
Note that the sequence, as discussed earlier, was designed to adopt two mutually exclusive helices.
We then applied our model to infer the structural ensembles from the data.

Similarly to what we have done in a previous paragraph, to challenge our model's ability to recover the correct structural ensemble from an inaccurate thermodynamic baseline ($F_0$ term in \autoref{eq:free_energy_maintext}), we performed the fitting procedure using two different baseline models.
We first used the default Turner 2004 parameters \citep{mathewsIncorporatingChemicalModification2004}, which were employed in the original sequence design.
We then repeated the analysis with the Andronescu 2007 parameters \citep{andronescuEfficientParameterEstimation2007}, which predict a substantially different initial ensemble with populations of  approximately 20\% helix 1 and 80\% helix 2, in stark contrast to the roughly 50-50 distribution predicted by the Turner 2004 parameters, as illustrated in Fig.~\ref{fig:fig_newseq}b, dotted lines.

The results of our model fitting are presented in the same panel as solid lines.
Remarkably, the two predictions of MERGE-RNA are consistent with each other, regardless of the initial thermodynamic parameter used, signaling robustness of the prediction with respect to the baseline parameters.
In both cases, the model adjusts the ensemble of structures so that the resulting ensemble slightly favors helix 1, in accordance with the lower mutation rates observed in the experimental data (Fig.~\ref{fig:fig_newseq}a).

Crucially, MERGE-RNA does more than adjust the populations of the two competing helices.
It reveals a gradient in base-pairing probabilities within regions A1 and A2, where nucleotides closer to the central domain (B) are more likely to be paired, reflecting the slope observed in the experimental mutation profiles (Fig.~\ref{fig:fig_newseq}a).
Beyond the two fully-formed helices, the model resolves intermediate states in which B is partially paired with both A1 and A2, as schematically illustrated in Fig.~\ref{fig:fig_newseq}c.
This mechanism, known as strand displacement (or strand invasion) \citep{ratajczykControllingDNARNA2025}, has been observed in biological systems and is consistent with the observed mutation profile: regions nearest the central domain are expected to pair first during displacement, and accordingly show reduced reactivity closer to the loops (Fig.~\ref{fig:fig_newseq}a).
The presence of such intermediate states is expected if the two loops are sufficiently stable.
Analysis of loop co-occupancy (\siautoref{tab:loop_cooccupancy}) demonstrates capture of intermediate states missed by standard models.
While baseline ensembles rarely allow both loops to form ($<0.2$\%), the model predicts simultaneous pairing in 42--57\% of structures, consistent with the observed mutation profiles and reflecting the strand-displacement mechanism in which the central domain transiently pairs with both competing helices.
We repeated the same analysis with the optimized Deigan pseudo-free-energy approach (Fig.~\ref{fig:designed_deigan_2x2}). Although the method partially captures the strand-displacement behavior, the inferred ensemble depends strongly on the probing concentration and on the baseline ensemble.
Overall, these results demonstrate MERGE-RNA's ability to capture a continuous spectrum of low-population intermediates that standard thermodynamic models fail to predict, independently of the baseline thermodynamic model.

\section*{Discussion}

In this work, we have introduced MERGE-RNA, a physics-based framework for the analysis of chemical probing data that addresses fundamental limitations of conventional approaches.
Our approach moves beyond the use of experimental data as simple pseudo-energetic restraints
\citep{deiganAccurateSHAPEdirectedRNA2009}
and instead models the entire experimental pipeline, from the physical interaction of the probe with the RNA to the final mutational readout.
This enables the determination of transferable, physically meaningful parameters and the integration of multiple datasets for inferring the native structural ensemble.
Like all current thermodynamics-based ensemble prediction methods, MERGE-RNA ultimately relies on the underlying nearest-neighbor thermodynamic model as its prior. The inferred soft constraints compensate for systematic discrepancies between this prior and the experimental data, but they cannot fully eliminate limitations arising from an incomplete physical description of RNA folding. Improvements in thermodynamic models are therefore expected to translate directly into improved ensemble predictions within our framework.
Applied to a homogeneous set of structured RNAs probed across multiple DMS concentrations and replicates, the model yields physical parameters that transfer across sequences and produces ensembles that agree with deposited structures more closely than the standard pseudo-free energy approach.
On the same systems, the sweet-spot behavior under uniform soft constraints, the improved correlation with mutation rates, and the localized disagreements between deposited structures and probing signals all point to the same conclusion:
our results suggest that even RNAs typically regarded as well-folded may not be fully described by a single static structure under the conditions of the probing experiment.
A more direct test of the inferred heterogeneity is provided by the \textit{V. vulnificus} adenine riboswitch, where NMR offers independent ground truth on the co-existing conformers and where other deconvolution methods have been benchmarked; here MERGE-RNA recovers the resolved conformations
and reproduces the ligand-dependent population shifts, including a ligand-dependent structural midpoint in agreement with the NMR-derived $K_d$.
MERGE-RNA predicts a somewhat smaller ligand-induced increase in the P1 population than reported from the NMR and DANCE-MaP analyses.
However, we regard this discrepancy to be minor compared with the overall agreement on the biologically relevant conformational transition considering the following factors: (i) the probing experiment was performed on a longer construct, which may have altered the relative populations of the two conformers; (ii) NMR and DANCE-MaP assume the existence of only a few discrete states, while MERGE-RNA allows for a more heterogeneous ensemble; (iii) other differences in the experimental conditions and inherent uncertainties in both the NMR and MERGE-RNA populations.
Interestingly, the Deigan pseudo-free-energy approach yields a similar apparent $K_d$, but substantially over-stabilizes the P1 helix and fails to reproduce the accompanying destabilization of the Shine–Dalgarno sequence required for the regulatory conformational switch.
Together with the deconvolution results on synthetic and experimental bistable constructs, these findings show that the framework can quantitatively characterize conformational populations, including the intermediates that arise in processes such as strand displacement and co-transcriptional folding.

We note that our model explicitly includes all four nucleotides even though DMS exhibits structure-dependent reactivity primarily with A and C. Including the complete sequence leads to more robust estimates of the physical parameters, ultimately improving the accuracy of inferred structural ensembles.
In principle, it could be extended to model pair-dependent reactivity in U and G nucleotides so as to fully exploit information reported by four-base DMS \cite{mustoeRNABasepairingComplexity2019}.
The separation of RNA thermodynamics (baseline model and $\lambda_i$) from probe-specific physics (all of the other parameters) enables future multi-probe extensions.
One could analyze data from CMCT \cite{ehresmannProbingStructureRNAs1987}, SHAPE \cite{merinoRNAStructureAnalysis2005,weeksAdvancesRNAStructure2010,marinusNovelSHAPEReagent2021}, hydroxyl radical footprinting \cite{adilakshmiHydroxylRadicalFootprinting2006}, and ETS \cite{doudsNewReagentVivo2024} simultaneously.
IPANEMAP \cite{saaidiIPANEMAPIntegrativeProbing2020} showed improvements from multi-probe integration but relies on empirical conversions.
In our framework, each probe would be modeled with its own parameters while each RNA molecule would have specific  $\lambda_i$ values, leveraging orthogonal chemical information.

A central strength is the ability to avoid over-constraining: standard pseudo-free energy approaches \cite{deiganAccurateSHAPEdirectedRNA2009} modify the energy landscape globally, often altering already-correct regions.
Similarly to Ref.~\citep{washietlRNAFoldingSoft2012},
our framework infers minimal soft constraints, ensuring modifications arise only where data support deviation, and resulting in a lower loss (Fig.~\ref{fig:redmond}b).
In addition, our physics-based model naturally enables the combination of multiple datasets
in a single fitting process.
Among ensembles compatible with experiment, the maximum entropy principle chooses the one closest to the prior, and modifications of the ensemble are here regularized by bounding soft constraints.
Alternative regularization approaches might be explored \cite{hummerBayesianEnsembleRefinement2015, cesariUsingMaximumEntropy2018}.

Several existing methods also aim to extract information about RNA structural ensembles from chemical probing data, although they address related rather than identical inference problems.
Existing ensemble-deconvolution pipelines such as DREEM \citep{tomezskoDeterminationRNAStructural2020}, DRACO \citep{morandiGenomescaleDeconvolutionRNA2021}, and DANCE-MaP \cite{olsonDiscoveryLargescaleCellstateresponsive2022} cluster sequencing reads with multiple co-occurring modifications and assign each cluster to a discrete structure.
By capitalizing on per-read correlation patterns, these methods recover information about alternative conformations but at the cost of discarding the abundant reads with zero or single modifications and demanding deep coverage to resolve low-probability states.
Alternative ensemble fitting approaches include Rsample \citep{spasicModelingRNASecondary2018} and SLEQ \citep{liStatisticalModelingRNA2018}.
Rsample employs a sample-and-cluster approach, reweighting the Boltzmann ensemble through pseudo-free energy adjustments resembling our soft constraints, though without physics-grounded calibration, but rather adopting an empirical strategy in a non self-consistent fashion.
SLEQ uses a sample-and-select strategy rooted in the maximum parsimony principle to identify the sparsest set of structures explaining modification patterns, differing from our maximum entropy approach designed to minimally perturb the thermodynamic model.
Our framework calibrates against averaged mutational fractions, ensuring every read (even negatives) contributes signal in sparse-modification regimes.
This distinction is illustrated by the adenine riboswitch analysed in this work, for which we use exactly the DMS dataset originally introduced in the DANCE-MaP study \citep{olsonDiscoveryLargescaleCellstateresponsive2022}. Despite relying only on the averaged mutation profile rather than read-level co-occurrence information, MERGE-RNA recovers the same ligand-dependent structural rearrangement reported by DANCE-MaP.
The trade-off is loss of explicit per-read co-variation, though reads could in principle be integrated to fine-tune populations. In fact, our framework enables the estimation of the likelihood of generating each read, which could be used to assign reads to structures and further refine population estimates, though we leave this as a future extension.
Read-level clustering would struggle to distinguish multiple similar conformations and represent continuous ensembles with many intermediates (e.g., strand-displacement pathways).
Furthermore, alternative methods assume each read originates from a single structure, which may not hold for RNAs undergoing rapid dynamics.
Our framework models ensemble-averaged behavior directly, accommodating fast interconversions at equilibrium.
Consequently, our framework captures both discrete alternative structures and continuous ensembles enabling interpretation of dynamic processes.
Strikingly, even without individual reads, the model reconstructs correlations and measures minor substate populations, which would be missed in methods that enumerate a small number of representative structures.

A key feature is predicting the structural ensemble at any probe concentration, especially extrapolating to zero concentration for the native state.
In our analyses, base-pairing probability dependence on concentration was modest, introducing partial parameter under-determination (e.g., increasing $\mu$ mimics decreasing $p_{\mathrm{bind}}$).
This does not compromise robustness or transferability (Fig.~\ref{fig:redmond}a) and can be mitigated by widening the concentration range; Bayesian treatment could be used to quantify identifiability.
At the same time, the model is equipped to handle systems where probes induce significant non-linearities \cite{calonaciMolecularDynamicsSimulations2023} from rearrangement, denaturation, or saturation. Standard protocols recommend avoiding these effects by reaching the single-hit kinetics \cite{arnoldInvestigatingInterplayRNA2025}, which might be difficult or impossible for large RNAs, and intrinsically eliminates correlation information.
Explicitly modeling these concentration-dependent perturbations might enable a more faithful representation of native conformational landscapes.

MERGE-RNA is intended for detailed ensemble inference on selected RNAs rather than transcriptome-wide analysis. Its optimization procedure is naturally more computationally demanding than a single thermodynamic folding calculation because it repeatedly evaluates partition functions during likelihood optimization. In the current implementation, the systems studied here required from approximately 1--2 hours for the smaller RNAs (e.g. the adenine riboswitch and cspA) to about 1--2 days for the largest benchmark systems on a single workstation. Nevertheless, the framework scales well to datasets comprising multiple measurements of the same RNA or collections of RNAs acquired under comparable experimental conditions, where the global physical parameters can be inferred jointly. Extending the approach to transcriptome-wide applications will require more efficient optimization strategies. Such strategies are currently under development and are expected to substantially improve the scalability of the framework.
An additional extension concerns RNA-protein interactions that confound probing readouts.
Known contacts could be excluded from the loss and constrained against internal pairing; unknown interaction sites could be flagged by localizing discrepancies between predicted and observed profiles.

We note that validation of RNA structure prediction methods typically relies on reference structures from crystallography, cryo-EM, or covariance analysis, each with inherent biases.
Crystals often fail to form for conformationally heterogeneous molecules, selecting against dynamic ensembles.
Cryo-EM reconstructions typically use only a small fraction of recorded particles, with difficult-to-estimate effects on captured dynamics.
Covariance-based methods reflect evolutionarily conserved, biologically functional structures, which may represent only a subset of conformations observed \textit{in vitro}.
These considerations highlight the value of probing-based ensemble methods as a complementary approach to traditional structure determination.

We acknowledge important limitations.
The underlying ViennaRNA framework does not account for pseudoknots, critical for many RNAs.
More generally, secondary structure alone may not suffice to predict DMS reactivities, which can depend on 3D features \citep{deenalatthaCharacterizingRNA3D2026}.
Possible factors neglected here include solvent accessibility, backbone geometry, stacking context, conformational dynamics, tertiary contacts, and the local electronic environment of the reactive nitrogen.
 In particular, low reactivity should not always be interpreted as evidence of canonical base pairing, since nucleotides may also be protected by tertiary contacts, non-canonical interactions, solvent accessibility, or other structural features not represented in the current model. Consequently, the present implementation may attribute part of this protection to secondary structure, potentially overestimating localized structural heterogeneity in some regions.
Despite these limitations, we here show that MERGE-RNA can
be used in a wide range of contexts, systematically increasing the correlation between base-pairing probability and observed DMS reactivity and recapitulating
reference RNA conformational ensembles.
Finally, while physical parameters transfer across RNAs under constant conditions, transferability between varying conditions (temperature, buffer, probing duration) is not guaranteed; we recommend fitting similar-condition data together.
Broader applicability would require modeling parameter dependence on experimental variables.

\section{Data and code availability}

The MERGE-RNA source code is available at \url{https://github.com/giusSacco/MERGE-RNA}
and has been archived on Zenodo under DOI: 10.5281/zenodo.20059549.
The repository includes scripts and instructions to reproduce the fits, run the analyses reported here, and generate the plots.
Sequencing data for all experiments performed in this work have been deposited in the European Nucleotide Archive (ENA) under accession number PRJEB113043.

\section{Competing interests}
No competing interest is declared.

\section{Author contributions statement}

Gi.S., Gu.S., and G.B. conceived and designed the study.
Gi.S., Gu.S., and G.B. developed the theoretical modeling and statistical framework, interpreted the data and the results.
Gi.S. implemented MERGE-RNA, led the primary data analysis, processed sequencing data, performed model benchmarking, and prepared the figures.
G.B. performed exploratory analyses and contributed methodological input.
J.L. and R.P.S. designed and performed the DMS-MaP experiments.
G.B., Gu.S. and R.P.S. supervised the research.
All authors wrote the manuscript, revised it critically, and approved the final version.

\section{Acknowledgments}
We thank Catherine Eichhorn, Anthony Mustoe and Joseph Yesselman for useful discussions.
Anthony Mustoe is also acknowledged for reading a preliminary version of this manuscript and providing useful suggestions.
Giuseppe Sacco acknowledges financial support from the European Molecular Biology Organization (EMBO) through the EMBO Scientific Exchange Grant 11403.
RPS acknowledges the interdisciplinary Thematic Institute IMCBio+, as part of the ITI 2021–2028 program of the University of Strasbourg, CNRS and Inserm, IdEx Unistra (ANR-10-IDEX-0002), SFRI-STRAT'US (ANR 20-SFRI-0012), and EUR IMCBio (ANR-17-EURE-0023) under the framework of the French Investments of the France 2030 Program.
GB acknowledges support from the Italian Ministry of University and Research (MUR) through the Fondo Italiano per la Scienza (FIS 3), Grant No. FIS-2024-01745, RNAScale.

\bibliography{data-driven-structure-prediction}

\clearpage

\setcounter{page}{1}
\renewcommand{\thepage}{S\arabic{page}}
\section*{Supplementary Information}

\setcounter{figure}{0}
\setcounter{table}{0}
\renewcommand{\thefigure}{S\arabic{figure}}
\renewcommand{\thetable}{S\arabic{table}}
\subsection*{Physical Model}\label{app:physical_model}

The physical model underlying our analysis of RNA chemical probing experiments is illustrated in the left panel of \autoref{fig:1} and will be detailed in this section.
We derive and justify the expression for the free energy of a secondary structure $s$ in the presence of a chemical probe (as reported in the main text) which enables the prediction of the ensemble of structures, and explain our methodology for producing a prediction of the reactivity profiles at varying probe concentrations, which serve as the theoretical basis for fitting experimental data.

\subsubsection*{Secondary structure populations}

The initial step in our analysis involves modeling the RNA folding into an ensemble of secondary structures, while also accounting for interactions with chemical probes.
Our baseline is the prediction of the ViennaRNA package \citep{lorenzViennaRNAPackage202011}, specifically employing the \texttt{RNA.fold\_compound.subopt} method to obtain an ensemble of suboptimal structures $s$ with their associated free energies $F_0(s)$.

The maximum entropy formalism \citep{demartinoIntroductionMaximumEntropy2018} provides a principled framework for integrating experimental data while making minimal adjustments to the baseline predictions from the classical thermodynamic model, ensuring that the resulting distribution of observables gets consistent with experimental measurements.
Within this framework, the minimal perturbation to the baseline model that aligns with experimental data is achieved by introducing sequence-specific model parameters $\lambda_i$ --one for each nucleotide position-- that serve as Lagrange multipliers constraining the ensemble to match experimentally observed properties.
These soft constraints can be effectively interpreted as positional corrections of pairing energy, allowing us to capture sequence-dependent effects that may not be fully accounted for in the standard thermodynamic model.
This formalism allows us to seamlessly integrate experimental data into the thermodynamic framework, and the corrections will be applied only where there are discrepancies between the model and the data, while leaving the areas where there is agreement unchanged.
Importantly, the $\lambda_i$ parameters are sequence-specific but remain consistent across all concentrations and experimental replicates for the same sequence.

The population of each secondary structure $s$ is determined by its free energy, which comprises three components: the baseline free energy $E_0(s)$, sequence-specific soft constraints $\lambda_i$, and the contribution from probe interactions $\Delta F_{\text{[DMS]}}(s)$:
\begin{align}
    F(s) &= F_0(s) + \sum_{i \in \text{paired}(s)} \lambda_i + \Delta F_{\text{[DMS]}}(s)
    \label{eq:free_energy_as_sum_of_three_terms}
\end{align}
The last term explicitly accounts for interactions between chemical probes and RNA structures, which modify the free energy in a concentration-dependent manner, whose explicit form is derived below.
The concentration-dependent energy of the probe-RNA interaction is described by chemical potential $\mu$.
Note that once a chemical potential for a reference probe concentration is known, the chemical potential for any other concentration can be derived from the ratio of the experimental concentrations as $\mu = \mu_r + \mathrm{k_BT}\ln\left(\frac{[\text{DMS}]}{[\text{DMS}]_r}\right)$, where $[\text{DMS}]$ is the concentration of the probe in the experiment and $[\text{DMS}]_r$ and $\mu_r$ are respectively the reference concentration and its chemical potential.
Since for most chemical probes the interaction energy depends on local RNA structure (this is especially true for the most widely used SHAPE and DMS), we include an additional parameter $\Delta\mu_{\text{pairing}}$ that represents the energetic penalty for probe interactions with paired nucleotides, such that the effective binding energy for paired nucleotides becomes $\mu' \equiv \mu - \Delta\mu_{\text{pairing}}$.
This formulation allows us to model the effect of varying probe concentrations with only two parameters ($\mu_r$ and $\Delta\mu_{\text{pairing}}$) and thus to combine data from multiple experiments into a single, coherent framework.

Importantly, in what follows we consider only the binding partition function, i.e., the contribution to the total partition function arising from independent probe-RNA binding degrees of freedom at fixed secondary structure $s$.
Under the assumption of independent binding across sites, this factorizes as
\begin{align}
    Z_{\text{bind}}(s) &= \prod_{i:\, s_i=0} z_u \prod_{i:\, s_i=1} z_p \\
    &= z_u^{N_u(s)}\, z_p^{N_p(s)}
\end{align}
with single-site partition functions $z_u = 1 + \mathrm{e}^{\beta\mu}$ for unpaired sites and $z_p = 1 + \mathrm{e}^{\beta\mu'}$ for paired sites, and where $N_u(s)$ and $N_p(s)$ denote the numbers of unpaired and paired nucleotides in structure $s$.
The corresponding concentration-dependent contribution to the free energy (the last term in \autoref{eq:free_energy_as_sum_of_three_terms}) is then
\begin{align}
    \beta \Delta F_{\text{[DMS]}}(s) &= -\ln\big(Z_{\text{bind}}(s)\big) \\
    &=-N_u(s)\ln z_u -N_p(s)\ln z_p\\
    &=-N_u(s)\ln(1+\mathrm{e}^{\beta \mu})-N_p(s)\ln(1+\mathrm{e}^{\beta\mu'})\\
    &=-(N - N_p(s))\ln(1+\mathrm{e}^{\beta \mu}) - N_p(s)\ln(1+\mathrm{e}^{\beta\mu'})\\
    &=- N\ln(1+\mathrm{e}^{\beta \mu}) - N_p(s)\ln\left(\frac{1+\mathrm{e}^{\beta\mu'}}{1+\mathrm{e}^{\beta\mu}}\right)
    \label{eq:derivation_Fprobe}
\end{align}
where $N \equiv N_u(s)+N_p(s)$ is the sequence length.
The first term is structure-independent and does not contribute to the populations in the experiment, so it can be safely ignored in our calculations.
In practice, this formulation enables us to account for the probe influence on the secondary structure populations by adding $-\ln\left(\frac{1+\mathrm{e}^{\beta\mu'}}{1+\mathrm{e}^{\beta\mu}}\right)$ as a soft constraint in ViennaRNA for each paired base, allowing us to efficiently compute pairing probabilities within its computational framework.
By substituting the expression for $\Delta F_{\text{[DMS]}}(s)$ into \autoref{eq:free_energy_as_sum_of_three_terms}, we obtain the final expression for the free energy of a secondary structure
\begin{align}
    F(s) &= \underbrace{F_0(s)}_{\text{baseline}} + \underbrace{\sum_{i \in \text{paired}(s)} \lambda_i}_{\substack{\text{site-specific}\\ \text{corrections}}} - \underbrace {\mathrm{k_B} T N_p(s)\ln\left(\frac{1+\mathrm{e}^{\beta\mu'}}{1+\mathrm{e}^{\beta\mu}}\right)}_{\substack{\text{concentration} \\ \text{dependent perturbation}}}
    \label{eq:free_energy_final}
\end{align}

\subsubsection*{Physical Binding Probability}\label{app:physical_binding}
Once we have outlined how to compute the populations of secondary structures in the presence of a chemical probe, we can derive the expected reactivity profile through the modeling of the chemical probing experiment, beginning from the interaction between the probe and the RNA.
The probability of physical binding between the probe and RNA follows a Boltzmann distribution:
\begin{align}
    \mathbb{P}(k_i=1|s_i=0) &= \frac{\mathrm{e}^{\beta\mu}}{1 + \mathrm{e}^{\beta\mu}}\\
    \mathbb{P}(k_i=1|s_i=1) &= \frac{\mathrm{e}^{\beta\mu'}}{1 + \mathrm{e}^{\beta\mu'}}
\end{align}
where $k_i \in \{0,1\}$ denotes the physical binding state at position $i$ (1 for probe-bound, 0 for unbound) and $s_i \in \{0,1\}$ represents the pairing state (1 for paired, 0 for unpaired).

We make two key assumptions in this model: first, that the probe physical interaction is independent of nucleotide identity, and second, that binding events occur independently at each site, with no interactions between probes.
Note that the latter may not hold at high probe concentrations, where cooperative effects and non-linear increases in reactivity can occur, as investigated by \citep{calonaciMolecularDynamicsSimulations2023} and \citep{arnoldInvestigatingInterplayRNA2025}.

\subsubsection*{Probability of Chemical Modification}
The probability of chemical modification at position $i$ depends on both the physical binding event and the subsequent conversion from physical to chemical modification.
We assume that when a probe physically interacts with nucleobase $i$, chemical modification occurs with a probability $p_{\text{bind}}(n_i, s_i)$, which depends only on the nucleotide identity $n_i \in \{\text{A, U, G, C}\}$ and its pairing state $s_i$:
\begin{align}
\mathbb{P}(c_i = 1| s_i, n_i)&= \mathbb{P}(c_i=1|k_i=1, s_i, n_i) \mathbb{P}(k_i=1|s_i) \\
&\equiv p_{\text{bind}}(n_i, s_i)\mathbb{P}(k_i=1|s_i)
\end{align}
where $c_i \in \{0,1\}$ represents the modification state (1 for chemically modified, 0 for unmodified).
While this is the most general formulation and holds for widely used chemical probes such as SHAPE, we note that DMS can react only with unpaired A and C and in a structure independent manner with G and U.
We can thus simplify the model and embed the physics of the probe in the model by setting
\begin{align*}
    p_{\text{bind}}(\text{A},1) &= 0 \\
    p_{\text{bind}}(\text{C},1) &= 0 \\
    p_{\text{bind}}(\text{G},0) &= p_{\text{bind}}(\text{G},1) \\
    p_{\text{bind}}(\text{U},0) &= p_{\text{bind}}(\text{U},1)
\end{align*}
This leads to a total of four free parameters for the nucleotide-specific binding probabilities, one for each nucleotide type.

The overall probability of chemical modification at site $i$, denoted $\mathbb{P}(c_i)$, is obtained by averaging over the structural ensemble.
Thanks to the assumption that binding and modification events are independent at each site, this average can be computed using the local pairing probabilities $\mathbb{P}(s_i)$, which are derived from the partition function without enumerating all possible secondary structures.
The ensemble-averaged probability of chemical modification is thus given by:
\begin{align}
    \mathbb{P}(c_i) &= \left\langle \mathbb{P}(c_i|s, n_i)\right\rangle_{s} \nonumber\\
    &= \left\langle \mathbb{P}(c_i|s_i, n_i)\right\rangle_{s_i} \nonumber\\
    &= \sum_{s_i \in \{0,1\}} \mathbb{P}(s_i)\mathbb{P}(c_i|s_i,n_i)
\end{align}

\subsubsection*{Expected Mutation Rate and Systematic Biases}\label{app:mut_rate_and_biases}

In Mutation Profiling (MaP), chemical modifications induced by probes are detected as mutations in the cDNA during reverse transcription. To accurately model this process, we must establish the relationship between chemical modifications and the observed mutations while accounting for errors introduced during reverse transcription.

Although mutation probabilities could potentially depend on both sequence context and structural environment \citep{verwiltArtifactsBiasesReverse2023}, we simplify our model by considering only the dependence on the chemical modification state $c_i$, treating other factors as systematic biases (which will be accounted for later in this section).
We introduce two parameters $m(c_i=1)$ and $m(c_i=0)$ that govern respectively mutation probabilities for modified and unmodified nucleotides, according to the following definition:
\begin{equation}
    \mathbb{P}(\text{mutation at position}~i~|~c_i) \equiv 1 - \mathrm{e}^{-(m(c_i)+\epsilon_i)}
\end{equation}

In this formulation, $m(c_i = 1)$ and $m(c_i = 0)$ can be interpreted as sequence-independent probabilities governing respectively false negative and false positive mutations and will be fitted as model parameters.
In the following and in the main text we use the shorthand notation $m_1 \equiv m(c_i=1)$ and $m_0 \equiv m(c_i=0)$.
The position-specific $\epsilon_i$ capture systematic biases common across all experimental replicates and probe concentrations for the same RNA sequence.
These $\epsilon_i$ values are calibrated to ensure that the model-predicted reactivity profile at zero probe concentration aligns with the control experiment, effectively serving as site-specific offsets common to all experimental conditions.

To compute the model prediction of the experimental reactivity profile, we average over all possible modification states, weighted by their probabilities:
\begin{align}
M_i &\equiv \langle M(c_i) \rangle_{c_i} \\
&= \sum_{j \in \{0,1\}} \mathbb{P}(c_i=j) M(c_i=j) \\
&= \sum_{j \in \{0,1\}} \mathbb{P}(c_i=j) (1 - \mathrm{e}^{-(m(j) + \epsilon_i)})\\
&= 1 - \mathrm{e}^{-\epsilon_i} \sum_{j \in \{0,1\}} \mathbb{P}(c_i=j) \mathrm{e}^{-m_j} \\
&= 1 - \mathrm{e}^{-\epsilon_i} \left[ (1 - \mathbb{P}(c_i=1)) \mathrm{e}^{-m_0} + \mathbb{P}(c_i=1) \mathrm{e}^{-m_1} \right] \\
&= 1 - \mathrm{e}^{-\epsilon_i} \left[ \mathrm{e}^{-m_0} + \mathbb{P}(c_i=1) (\mathrm{e}^{-m_1} - \mathrm{e}^{-m_0}) \right]
\end{align}

This equation establishes a critical link between experimentally observed mutation rates and the underlying RNA structural ensemble through the chemical modification probabilities $\mathbb{P}(c_i=1)$, which themselves depend on the pairing state probabilities $\mathbb{P}(s_i)$ derived earlier.

Notably, the first term in the square brackets, containing $m_0$, produces a common background signal. When control experiments are available, this term can be incorporated into the systematic biases $\epsilon_i$; otherwise, it must be fitted as a constant background term.

\subsubsection*{Model Parameters Summary}

Before moving on to how to obtain the optimal model parameters from experimental data, we summarize here all the parameters of our model that require fitting:
\begin{itemize}
    \item \textbf{Physical parameters}, which describe the physics of the chemical probing experiment. These are shared across all RNA sequences and probe concentrations.
    \begin{itemize}
        \item $\mu_r$: quantifies the energy of probe-RNA interaction at a reference probe concentration, from which we derive chemical potentials for all other concentrations.
        \item $\Delta\mu_{\text{pairing}}$: represents the energetic penalty for probe interaction with paired nucleobases.
        \item $p_{\text{bind}}(n_i, s_i)$: is the probability of chemical modification given nucleotide identity $n_i$ and pairing state $s_i$. These are four for DMS, one for each nucleotide type.
        \item $m_1$ and $m_0$: are parameters governing respectively false negative and false positive mutation probabilities during the cDNA transcription. If control experiments are available, $m_0$ gets absorbed into the systematic biases $\epsilon_i$ and does not need to be fitted.
    \end{itemize}
    \item \textbf{Soft constraints}, which are sequence-specific but consistent across all experimental conditions for a given RNA:
    \begin{itemize}
        \item $\lambda_i$: are site-specific soft constraints modifying the pairing energy of the baseline thermodynamic model.
    \end{itemize}
\end{itemize}

Worth mentioning in this summary--although not fitted--are the $\epsilon_i$, which are calibrated from the control experiment and incorporate site-specific systematic errors shared across replicates and concentrations, and effectively serve as offsets to the predicted mutation profiles.
They enable the decoupling of experimental biases from the thermodynamic and concentration-dependent effects, allowing the model to focus on physically relevant signal.

\paragraph*{\textit{Computation of base-pairing probabilities}}
All base-pairing probabilities (BPPs) reported in this work were computed with ViennaRNA v2.7.0 \citep{lorenzViennaRNAPackage202011} through its Python interface, using the default Turner 2004 nearest-neighbour parameters \citep{turnerNNDBNearestNeighbor2010} unless otherwise stated; results with the Andronescu 2007 parameters \citep{andronescuEfficientParameterEstimation2007}, loaded via \texttt{RNA.params\_load\_RNA\_Andronescu2007()}, are reported alongside where indicated. For each sequence the folding temperature is set with (\texttt{RNA.cvar.temperature}) and we build a \texttt{fold\_compound}; any per-nucleotide energy correction is added to this \texttt{fold\_compound} first, and only then do we evaluate the partition function (\texttt{fc.pf()}) and read the equilibrium BPP matrix (\texttt{fc.bpp()}). The three models compared in \autoref{fig:redmond} share this identical pipeline and differ only in the correction applied before \texttt{fc.pf()}:
\begin{itemize}
    \item \textbf{ViennaRNA baseline} (reactivity-agnostic): no correction; the unmodified thermodynamic ensemble.
    \item \textbf{Deigan pseudo-free-energy} \citep{deiganAccurateSHAPEdirectedRNA2009}: after subtraction of the control and clipping of negatives to zero, reactivities are normalised by the standard box-plot procedure (values above the $Q_3+1.5\,\mathrm{IQR}$ outlier fence are discarded and the profile is divided by the mean of the top $10\%$ of the remaining values), computed over the probed adenine and cytosine positions only, then converted to a per-position pseudo-free-energy through ViennaRNA's dedicated \texttt{fc.sc\_add\_SHAPE\_deigan(reactivities, m, b)}, applied to adenines and cytosines only. Rather than adopting the literature defaults $(m,b)=(1.8,-0.6)$, the slope and intercept were refit to our dataset, as detailed in the SI Appendix (\nameref{si:deigan_refit}).
    \item \textbf{MERGE-RNA}: during the fit we apply, on top of the baseline, both the inferred site-specific soft constraints $\lambda_i$ and the concentration-dependent penalty on paired bases $\Delta F_{\text{[DMS]}}$ (\autoref{eq:free_energy_maintext}), which acts as a uniform free-energy penalty applied to every paired nucleotide. Both enter through the same ViennaRNA soft-constraint interface: a per-nucleotide energy can be added equivalently either to a base's unpaired state (\texttt{sc\_add\_up}) or to all of its base pairs (\texttt{sc\_add\_bp}), the two yielding the same normalized ensemble, and we choose between them for numerical stability (for the $\lambda_i$, according to their sign). The extrapolated BPPs at zero probe concentration are then obtained by setting $\Delta F_{\text{[DMS]}}=0$.
    To optimize the model parameters, derivatives of the base-pairing probabilities with respect to the optimized parameters are computed from conditional pairing probabilities obtained by imposing hard constraints that force individual nucleotides to remain unpaired (\texttt{hc\_add\_up}).
\end{itemize}
In all cases the corrections enter the ViennaRNA recursions directly, so the partition function is recomputed with the modified energies and the BPPs reflect the full, exactly re-weighted Boltzmann ensemble. We do not enumerate or re-score a finite set of suboptimal structures a posteriori. Should individual representative structures be of interest, they can be obtained downstream from the same constrained ensemble (e.g.\ by stochastic backtracking or \texttt{RNA.fold\_compound.subopt}); while this can be useful for downstream analyses, it plays no role in the fit.

\subsection*{Model Training}\label{app:model_fitting}

\subsubsection*{Loss function}\label{app:loss_function}
To fit the model parameters to experimental data, we define a loss function that quantifies the discrepancy between the expected reactivity profiles predicted by our model (as derived in the previous section) and the observed mutation rates from chemical probing experiments.

The binomial nature of the mutation counts, which arise from a series of independent Bernoulli trials (each read can either show a mutation or not), motivates the choice of a binomial likelihood for our loss function.
Within this framework, the probability of observing $\mathcal{M}_i$ mutations out of a coverage of $n_i$ reads under predicted mutation rate $M_i$ is given by:
\begin{align}
    \mathbb{P}(\mathcal{M}_i) &= \text{Binomial}(\mathcal{M}_i, n_i, M_i) \\
    &= \begin{pmatrix} n_i \\ \mathcal{M}_i \end{pmatrix}  M_i^{\mathcal{M}_i} (1-M_i)^{n_i - \mathcal{M}_i}
\end{align}
where $i$ runs over all the data points, i.e. sequences, positions, concentrations and replicates.
We thus define the loss function as the negative log-likelihood of observing the experimental mutation counts $\mathcal{M}_i^{\text{exp}}$ given the model.
\begin{align}
   \mathcal{L} &= \sum_i - \ln \mathbb{P}(\mathcal{M}_i^{\text{exp}} = \mathcal{M}_i) \\
   &= \sum_i - \ln \left[\begin{pmatrix} n_i \\ \mathcal{M}_i^{\text{exp}} \end{pmatrix} M_i^{\mathcal{M}_i^{\text{exp}}} (1-M_i)^{n_i - \mathcal{M}_i^{\text{exp}}}\right]
   \label{eq:loss}
\end{align}

\subsubsection*{Fit of parameters}\label{app:fit_parameters}

The model fitting process involves optimizing a set of parameters to best reproduce experimental data.
We implemented the optimization procedure in a custom Python script utilizing the \texttt{scipy.optimize.minimize} function \citep{virtanenSciPy10Fundamental2020} with the L-BFGS-B algorithm \citep{byrdLimitedMemoryAlgorithm1995}.
This limited-memory quasi-Newton method is particularly well-suited for high-dimensional optimization problems, as it efficiently approximates the inverse Hessian matrix while maintaining reasonable memory requirements.
The algorithm's ability to handle bound constraints is especially valuable for ensuring physically meaningful parameter values throughout the optimization process.
Specifically, we constrained the parameters to physically plausible ranges: the chemical potential at reference concentration, $\mu_r$, was bounded between -5 and 5 kcal/mol; the energetic penalty for pairing, $\Delta\mu_{\text{pairing}}$, was constrained to be non-negative, with an upper bound of 10 kcal/mol; the chemical modification probabilities, $p_{\text{bind}}$, were constrained to the range $[0, 1]$, as is required for probabilities; the false positive mutation parameter, $m_0$, was kept small within $[0, 0.1]$; and the false negative parameter, $m_1$, was bounded within $[0.1, 10]$.

The optimization is performed in a two-stage process to minimize the loss function.
In the first stage, we fit the physical parameters that are common to all experiments, keeping the sequence-specific soft constraints ($\lambda_i$) fixed.
This global optimization starts from random initializations and runs until convergence.
In the second stage, the optimized physical parameters are held constant while we fit the $\lambda_i$ soft constraints independently for each RNA sequence.
This hierarchical strategy allows us to first establish a robust physical model and then refine the secondary structure populations with sequence-specific adjustments.
During the optimization, the $\lambda_i$ parameters are constrained to the range $[-1, 1]$ kcal/mol to ensure proper regularization and avoid overfitting.

\subsection*{Experimental Procedures}

\subsubsection*{In vitro transcription (IVT)}
A single-stranded full-length DNA oligonucleotide (\dna{5'-ATGGTCTGCTGGAGGTTGGAATAGTTGTGAGTTGAAGTGGGGATTTTTTAATCCCCACTTCAACTCACAACTATTCCAATTTTTTGGAATAGTTGTGAGTTGAAGTGGGGATTGGCTGTGGCATAATG-3'}) was chemically synthesized (Integrated DNA Technologies). This full-length sequence was amplified by PCR using primers carrying a T7 promoter. The primers used were Full-T7-RV (\dna{5'-TAATACGACTCACTATAGGGCATTATGCCACAGCCAATCCCCACTTC-3'}) and Full-FW (\dna{5'-ATGGTCTGCTGGAGGTTGGAATAGTTGTGAG-3'}).

PCR was performed in 50 \textmu L reactions containing the following final concentrations: GXL reaction buffer (1$\times$, from a 5$\times$ stock; Takara Bio), each dNTP (0.2 mM), each primer (0.25 \textmu M; Full-T7-RV and Full-FW), single-stranded full-length DNA template (1 nM), and PrimeSTAR GXL DNA Polymerase (0.025 U/\textmu L; Takara Bio). Thermal cycling conditions were as follows: initial denaturation at 98 $^\circ$C for 2 min; 25 cycles of 98 $^\circ$C for 20 s, 55 $^\circ$C for 20 s, and 68 $^\circ$C for 30 s; followed by a final extension at 68 $^\circ$C for 5 min. PCR products were analyzed on a 2\% agarose gel to verify the expected size, before purification using the NucleoSpin Gel and PCR Clean-up kit with NTC buffer (Macherey--Nagel), according to the manufacturer's instructions. The purified DNA was used as a template for T7 RNA polymerase\,\textemdash\,mediated in vitro transcription (IVT). IVT was performed in 50 \textmu L reactions containing the following final concentrations: T7 RNA Polymerase Reaction Buffer (1$\times$, from a 10$\times$ stock; New England Biolabs), ATP, CTP, GTP and UTP (each 5 mM; Ribonucleotide Solution Mix, New England Biolabs), RNase Inhibitor, Murine (1 U/\textmu L; New England Biolabs), T7 RNA Polymerase (2.5 U/\textmu L; New England Biolabs), DNA template (25 nM), and inorganic pyrophosphatase (0.0005 U/\textmu L; Thermo Scientific), in nuclease-free water. Reactions were incubated at 37 $^\circ$C for 2 h. Positive IVT control reactions were set up identically, except that a previously validated DNA template was used. After IVT, residual DNA template was removed by DNase digestion. DNase digestion was carried out at 37 $^\circ$C for 30 min in reactions containing the following final concentrations: DNase TURBO (0.1 U/\textmu L) and DNase TURBO buffer (1$\times$; Invitrogen). IVT products with and without DNase treatment were analyzed on a 2\% agarose gel to assess the efficiency of DNA removal and the quality of the RNA transcripts.

\subsubsection*{RNA refolding}
DNase-treated IVT products were column-purified and used for RNA refolding. For the initial refolding step, RNA was diluted in ultrapure water to an RNA concentration of 26 nM, and refolding buffer was added to reach the following final concentrations: EDTA (0.5 mM), HEPES (200 mM, pH 7.5), NaCl (300 mM). The mixture was heated at 95 $^\circ$C for 1.5 min to denature RNA secondary structures and then immediately placed on ice. While on ice, MgCl$_2$ was added to give a final concentration of 5 mM and an RNA concentration of 23.5 nM. Samples were then incubated at 37 $^\circ$C for 30 min to allow RNA refolding. Refolded RNA was stored at $-80$ $^\circ$C until further use.

\subsubsection*{DMS probing}
For dimethyl sulfate (DMS) probing, reactions were prepared with refolded RNA at a final concentration of 23.5 nM. DMS (Sigma--Aldrich) stock solutions were prepared in 100\% ethanol such that, upon addition to the RNA samples, the final DMS concentrations were 20, 40, 60, 80, or 100 mM. $\beta$-mercaptoethanol (Sigma--Aldrich) was diluted in PBS such that, upon addition to the reaction, its final concentration was 1000 mM. For each DMS concentration, one negative control (receiving an equal volume of 100\% ethanol instead of DMS) and two technical replicates with DMS were prepared. Refolded RNA samples were kept on ice, mixed thoroughly with the appropriate DMS working solution or with 100\% ethanol for negative controls, and immediately transferred to an incubator pre-equilibrated to 37 $^\circ$C for 7 min. Reactions were then placed back on ice, and $\beta$-mercaptoethanol working solution was added to each reaction and mixed thoroughly to reach its final concentration of 1000 mM and to quench DMS modification.

\subsubsection*{Reverse transcription}
DMS-treated RNA was column-purified and subjected to reverse transcription using MarathonRT \citep{guoSequencingStructureProbing2020,bohnNanoDMSMaPAllowsIsoformspecific2023,gribling-burrerIsoformspecificRNAStructure2024}. Primer\,\textendash\,RNA annealing reactions were prepared on ice with the following final concentrations: each dNTP (0.5 mM) and Full-FW primer (0.25 \textmu M). Primer\,\textendash\,RNA mixtures were incubated at 65 $^\circ$C for 5 min and then rapidly cooled on ice. Reverse transcription was then performed in 40 \textmu L reactions containing the primer\,\textendash\,RNA annealing mixture and the following final concentrations: MarathonRT reaction buffer (1$\times$; final composition 50 mM Tris\,\textendash\,HCl, pH 8.3; 200 mM KCl; 20\% (v/v) glycerol, prepared from a 3$\times$ stock containing 150 mM Tris\,\textendash\,HCl, pH 8.3; 600 mM KCl; 60\% (v/v) glycerol), MnCl$_2$ (1 mM), MarathonRT (1 U/\textmu L; in-house purified, Addgene plasmid no. 109029 \citep{zhaoUltraprocessiveAccurateReverse2018}; http://n2t.net/addgene:109029; RRID: Addgene\_109029; stored in 50 mM Tris\,\textendash\,HCl, pH 8.3; 200 mM KCl; 20\% (v/v) glycerol), RNase Inhibitor, Murine (1 U/\textmu L; New England Biolabs), and DTT (5 mM). No\,\textendash\,reverse transcriptase (no\,\textendash\,RT) controls were prepared in parallel with the same components and final concentrations, except that MarathonRT was omitted.

\subsubsection*{PCR amplification}
cDNA products were column-purified and subsequently amplified by PCR. PCR amplification was performed in 50 \textmu L reactions containing the following final concentrations: GXL reaction buffer (1$\times$, from a 5$\times$ stock; Takara Bio), each dNTP (0.2 mM), each primer (0.25 \textmu M; Full-T7-RV and Full-FW), and PrimeSTAR GXL DNA Polymerase (0.025 U/\textmu L; Takara Bio). An appropriate amount of cDNA was used as template. Thermal cycling conditions were: 98 $^\circ$C for 2 min; 25 cycles of 98 $^\circ$C for 20 s, 55 $^\circ$C for 20 s, and 68 $^\circ$C for 30 s; followed by a final extension at 68 $^\circ$C for 5 min. PCR amplicons were analyzed on a 2\% agarose gel and stained with ethidium bromide (EtBr) to assess amplification quality.

\subsubsection*{Nanopore sequencing}
PCR amplicons were column-purified, and DNA concentrations were determined using the Qubit dsDNA HS Assay Kit (Thermo Fisher Scientific). After normalizing the molar concentration of each sample, a total of 1{,}600 fmol DNA was used in library preparation, with 5 \textmu L taken from each sample. For each sample, dA-tailing and 5$'$ phosphorylation were performed in 6 \textmu L reactions containing 5 \textmu L DNA, 0.7 \textmu L NEBNext End-Repair Buffer (1$\times$ final concentration; New England Biolabs), and 0.3 \textmu L NEBNext End-Repair Enzyme Mix (New England Biolabs). Reactions were incubated at 20 $^\circ$C for 10 min, followed by 65 $^\circ$C for 10 min. Barcoding ligation was performed using the SQK\,\textendash\,NBD114\,\textendash\,96 kit (Oxford Nanopore Technologies). For each sample, barcoding ligation was carried out in 5 \textmu L reactions containing 1.5 \textmu L end-repaired DNA, 1 \textmu L barcode, and 2.5 \textmu L NEB Blunt/TA Ligase Master Mix (New England Biolabs, added as supplied). Reactions were incubated at room temperature for 20 min, after which ligation was quenched by adding 1 \textmu L EDTA (SQK\,\textendash\,NBD114\,\textendash\,96). Subsequently, 5 \textmu L of each barcoded sample was pooled, and the resulting pool was purified with 1$\times$ AMPure XP beads (SQK\,\textendash\,NBD114\,\textendash\,96) and washed twice with Short Fragment Buffer (SFB; SQK\,\textendash\,NBD114\,\textendash\,96). The pooled barcoded DNA was eluted in 35 \textmu L nuclease-free water. Adapter and motor protein ligation was performed in 56 \textmu L reactions containing 35 \textmu L pooled barcoded DNA, 11 \textmu L NEBNext Quick Ligation Reaction Buffer (New England Biolabs, B6058S; 1$\times$ final concentration), 5 \textmu L Native Adapter (NA; SQK\,\textendash\,NBD114\,\textendash\,96, Oxford Nanopore Technologies), and 5 \textmu L high-concentration NEB T4 DNA Ligase (New England Biolabs, T2020M). Reactions were incubated at room temperature for 20 min. The library was then purified with 0.6$\times$ AMPure XP beads (SQK\,\textendash\,NBD114\,\textendash\,96) and washed twice with SFB, taking care to avoid drying of the beads during washing and prior to elution. The final library was loaded onto an R10.4.1 flow cell (FLO\,\textendash\,PRO114M; Oxford Nanopore Technologies) and sequenced on a PromethION 2 Solo instrument (Oxford Nanopore Technologies) controlled by MinKNOW software (version 25.09.16). Basecalling was performed with Dorado v1.1.1 (Oxford Nanopore Technologies) in super-accuracy (``sup'') mode using the \texttt{dna\_r10.4.1\_e8.2\_400bps\_sup@v5.2.0} model.

\subsection*{Mapping and counting of mutations}

Sequencing reads were processed with an RNAFramework-based \citep{incarnatoRNAFrameworkAllinone2018} workflow using the functions \texttt{rf-map} and \texttt{rf-count}, following a similar pipeline to \citep{morandiGenomescaleDeconvolutionRNA2021} with minor adaptations for our constructs.
For each library, we generated a Bowtie2 \citep{langmeadFastGappedreadAlignment2012} index from the reference sequence using \texttt{bowtie2-build} and produced the companion FASTA index with \texttt{samtools faidx} \citep{danecekTwelveYearsSAMtools2021}.
Reads were aligned to the indexed reference with \texttt{rf-map -cq5 20 -cqo -mp '--very-sensitive-local' -b2 -bi ../\{ref\_fasta\}\_index}, where \texttt{\{ref\_fasta\}\_index} are the files produced by \texttt{bowtie2-build}.
The resulting BAM files were coordinate sorted and indexed (\texttt{samtools sort/index}) before quantification.

Mutation profiles were obtained with \texttt{rf-count} \cite{incarnatoRNAFrameworkAllinone2018} using the options recommended by Ref.~\citep{morandiGenomescaleDeconvolutionRNA2021} (\texttt{-m -ds 75 -na -ni -md 3}).These parameters report mutational load rather than RT stops (\texttt{-m}), discard reads shorter than 75 nt (\texttt{-ds 75}), ignore ambiguous deletions (\texttt{-na}) and indels (\texttt{-ni}), and cap deletions at three nucleotides (\texttt{-md 3}). The full command sequence executed for each FASTQ file is summarized below:

\begin{verbatim}
# build indices of reference
bowtie2-build --quiet {ref_fasta} {ref_fasta}_index
samtools faidx {ref_fasta}
# alignment (see above)
rf-map \
    -cq5 20 \
    -cqo \
    -mp '--very-sensitive-local' \
    -b2 \
    -bi {ref_fasta}_index \
    -o rf_map_{output_suffix} \
    {fastq_file}
# sorting and indexing the BAM file
samtools sort {bam_file} -o {sorted_bam_file}
samtools index {sorted_bam_file}
# count mutations to obtain mutation profile and mutation map
rf-count \
    -f {ref_fasta} \
    -m \
    -ds 75 \
    -na \
    -ni \
    -md 3 \
    {sorted_bam_file}
\end{verbatim}

For the reference structured RNAs (HCV IRES, bacterial RNase P, \textit{V. cholerae} glycine riboswitch, \textit{Tetrahymena} ribozyme, HC16), we instead adopted the filtering scheme described by the original work \citep{bohnNanoDMSMaPAllowsIsoformspecific2023}.
\begin{verbatim}
rf-count -r -p {num_threads} \
    -mf {primer_mask} \
    -f {reference_fasta} \
    --only-mut 'G>Y;A>B;C>D;T>V' \
    -m \
    -nd \
    -ni \
    -q {minq} \
    -eq 10
\end{verbatim}

All downstream structural analyses reported in the main text rely on the coverage and mutation rates generated by this standardized workflow.

\subsection*{Importance of conformational heterogeneity in structured RNAs}\label{si:heterogeneity}

\begin{figure}[h]
    \centering
    \includegraphics[width=.8\linewidth]{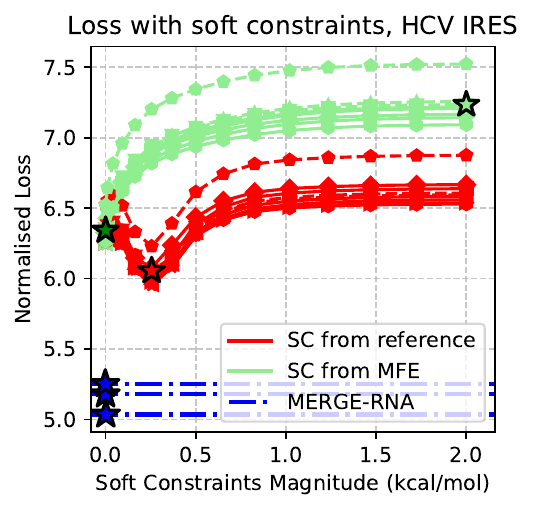}
    \caption{Effect of a uniform soft constraint on the agreement with chemical-probing data for HCV IRES.
    For each scalar magnitude on the x-axis, every nucleotide annotated as paired in the chosen reference structure is assigned that value as its soft constraint, all other positions are set to zero, and the physical parameters are kept fixed; this smoothly interpolates the ensemble between the original thermodynamic prediction and a single constrained structure.
    The y-axis reports the per-datapoint loss (\siautoref{eq:loss}).
    Red line: constraints derived from the experimental reference structure; light-green line: constraints derived from the minimum-free-energy (MFE) structure predicted by the thermodynamic model.
    Each curve corresponds to an independent fit in which the physical parameters were trained on a different triplet of RNA systems (cross-validation sets of Fig.~\ref{fig:redmond}a).
    The horizontal blue line indicates the loss reached by MERGE-RNA when the sequence-specific $\lambda_i$ are inferred position-by-position.
    The non-monotonic behavior of the red curve, together with the lower loss attained by the blue line, illustrates that explaining the observed mutation profiles requires retaining ensemble heterogeneity rather than collapsing onto any single structure, and that this is most effectively achieved by inferring per-nucleotide soft constraints.}
    \label{fig:fig2b_HCV_IRES}
\end{figure}

We found it instructive to illustrate, with a simple controlled experiment, why ensemble heterogeneity is needed to explain chemical-probing data and where the compromise between structural specificity and conformational diversity lies.
Using HCV IRES as a representative example, we applied a uniform soft constraint instead of inferring it position-by-position (Fig. \siref{fig:fig2b_HCV_IRES}, analogous results for other systems in Figs. \siref{fig:fig2b_tetrahymena_ribozyme},\siref{fig:fig2b_V_chol_gly_riboswitch},\siref{fig:fig2b_hc16}): for a given scalar magnitude (x-axis), every nucleotide annotated as paired in the chosen reference structure (experimental or minimum-free-energy) was assigned that value, all other positions were set to zero, and the physical parameters were kept fixed. Increasing this magnitude progressively biases the ensemble toward the targeted structure, smoothly interpolating between the original thermodynamic ensemble and a single constrained structure.

By interpolating between the original ensemble and its single most-likely structure (minimum-free-energy),
we assess the impact of reducing ensemble heterogeneity on the agreement with experiment (light green line in Fig.~\ref{fig:fig2b_HCV_IRES}).
In fact, favoring a single structure has the effect of progressively suppressing the population of alternative conformations that are likely present in solution and contribute to the observed mutation profiles.
Interestingly, the loss function monotonically increases when the
constraint is applied, degrading the agreement with experiment.
On the other hand, incorporating knowledge from
the reference experimental structure (red line in Fig.~\ref{fig:fig2b_HCV_IRES}) initially improves the model's agreement with experimental data.
However, even in this case, excessive constraint enforcement eventually degrades performance.
This pattern indicates the existence of a ``sweet spot'' where the model partially incorporates reference structural information while preserving the conformational diversity necessary to explain the experimental measurements.
Too strong a constraint toward any single structure—even the experimental reference—reduces the model's ability to account for the heterogeneous structural ensemble that is captured in the experimental data.
Overall, this analysis shows that a proper balance of information from the reference single experimental structure and ensemble diversity improves the capability of the model to reproduce chemical probing experiments.

\subsection*{Refitting the Deigan slope and intercept}\label{si:deigan_refit}

The Deigan pseudo-free-energy method \citep{deiganAccurateSHAPEdirectedRNA2009} enters our benchmark through exactly the same ViennaRNA folding pipeline as MERGE-RNA, described above under ``\textit{Computation of base-pairing probabilities}'': a reactivity $r_i$ is converted into the per-nucleotide restraint $\Delta G_i = m\,\ln(1+r_i)+b$ and added to the folding free energy with \texttt{sc\_add\_SHAPE\_deigan}, applied to adenines and cytosines only. Only two ingredients are specific to this baseline: the reactivities passed to the restraint and the coefficients $(m,b)$.

\paragraph*{\textit{Input reactivities.}}
We start from the same mutation-rate data used by MERGE-RNA and subtract the untreated control, clip negative values to zero.
The resulting reactivities are normalised following the standard box-plot normalisation scheme: values above the $Q_3+1.5\,\mathrm{IQR}$ box-plot outlier fence are discarded and the profile is divided by the mean of the top $10\%$ of the remaining values, before passing it to \texttt{sc\_add\_SHAPE\_deigan}. Because DMS probes only adenine and cytosine, this normalisation factor is computed over the A/C positions only. All downstream comparisons use the same per-position masking as the corresponding MERGE-RNA fits, so that the two methods see identical data.

\paragraph*{\textit{Selection of the coefficients $(m,b)$.}}
The slope and intercept are usually taken from the literature, but their optimum depends on the probe, the experimental conditions and the thermodynamic model, and is commonly re-optimized on the dataset at hand \citep{morandiGenomescaleDeconvolutionRNA2021}. We borrow the grid-search selection of RNAFramework's \texttt{rf-jackknife} \citep{incarnatoRNAFrameworkAllinone2018}: we scan $m\in[0,5]$ and $b\in[-3,0]$ with step $0.2$ ($416$ combinations) and, for each pair, fold every system with the Deigan restraint and score the predicted minimum-free-energy structure against the reference by the relaxed Fowlkes--Mallows index ($\mathrm{FMI}=\sqrt{\mathrm{PPV}\cdot\mathrm{sensitivity}}$, $\pm1$-nt tolerance, pseudoknots and lonely pairs removed), as previously done in \citep{morandiGenomescaleDeconvolutionRNA2021}. The $(m,b)$ pair with the highest geometric mean of the FMI over all systems is selected as the optimum.
For the structured RNAs of Fig.~\ref{fig:redmond} the reference structures are themselves the evaluation target, so choosing $(m,b)$ on them would be circular; we therefore select the coefficients by leave-one-system-out at each DMS concentration: for each held-out system, $(m,b)$ is optimized on the other three at the same concentration and applied to the held-out one, obtaining one coefficient pair per concentration and per held-out system.
The joint per-concentration optimum is given in Table~\ref{tab:deigan_mb_joint}; Table~\ref{tab:deigan_mb} reports instead, for every system and concentration, the per-held-out-system coefficients used for the Fig.~\ref{fig:redmond} bars and the two oracle performance ceilings (Frobenius and probing $R^2$, defined below). For the other RNAs analysed in this work no single reference structure is available, so the protocol above is not applicable. Since those experiments were all probed at $[\mathrm{DMS}]\ge 57$~mM, we reuse the $57$~mM grid optimum over the four PDB-referenced systems (without any leave-out) as a fixed, transferable coefficient, $(m,b)=(2.4,-0.4)$ under Turner and $(2.4,-1.0)$ under Andronescu.

\paragraph*{\textit{Performance ceiling.}}
In the main text the solid Deigan bars use the coefficients selected above (held-out for the structured RNAs, transferable for the others). The dashed caps instead report the \emph{ceiling} of the Deigan functional form: on the same $(m,b)$ grid we score each pair directly with the metric of interest: the Frobenius distance in Fig.~\ref{fig:redmond}b and the probing $R^2$ in Fig.~\ref{fig:redmond}d, in place of the FMI used for selection and keep its optimum, which is by construction the best value any choice of $(m,b)$ can attain for that metric, and therefore an oracle upper bound on what this two-parameter form can achieve.

\begin{table}[t]
\centering
\caption{Joint Deigan optimum: slope and intercept $(m,b)$ that maximize the geometric-mean
FMI over all four PDB-referenced systems at each DMS concentration (per-concentration, no
leave-out), under the Turner 2004 and Andronescu 2007 baselines. The per-system, held-out and
oracle-ceiling coefficients are reported in Table~\ref{tab:deigan_mb}; the literature default
is $(1.8,-0.6)$.}
\label{tab:deigan_mb_joint}
\begin{tabular}{ccc}
\toprule
$[\mathrm{DMS}]$ (mM) & Turner $(m,b)$ & Andronescu $(m,b)$ \\
\midrule
$8$ & $(0.0,\,\phantom{-}0.0)$ & $(1.4,\,-0.2)$ \\
$17$ & $(2.0,\,\phantom{-}0.0)$ & $(1.0,\,\phantom{-}0.0)$ \\
$34$ & $(1.8,\,\phantom{-}0.0)$ & $(4.0,\,-0.6)$ \\
$57$ & $(2.4,\,-0.4)$ & $(2.4,\,-1.0)$ \\
\bottomrule
\end{tabular}
\end{table}

\begin{table*}[p]
\centering
\caption{Deigan slope and intercept $(m,b)$ for every PDB-referenced system and DMS
concentration, under the Turner 2004 and Andronescu 2007 baselines, obtained on the
$m\in[0,5]$, $b\in[-3,0]$ grid (step $0.2$).
\emph{Held-out}: leave-one-system-out optimum used for the Fig.~\ref{fig:redmond} bars,
optimized on the other three systems at the same concentration. \emph{Ceiling$_{F}$} and
\emph{Ceiling$_{R^2}$}: oracle optima that instead minimize the Frobenius distance
(Fig.~\ref{fig:redmond}b) or maximize the probing $R^2$ (Fig.~\ref{fig:redmond}d) directly on
the same grid, i.e.\ the best value the two-parameter form can attain for that metric. The
joint per-concentration optimum is reported separately in Table~\ref{tab:deigan_mb_joint};
the literature default is $(m,b)=(1.8,-0.6)$.}
\label{tab:deigan_mb}
\begin{tabular}{@{}l r ccc@{}}
\toprule
\multirow{2}{*}{System} & $[\mathrm{DMS}]$ & Held-out & Ceiling$_{F}$ & Ceiling$_{R^2}$ \\
 & (mM) & $(m,b)$ & $(m,b)$ & $(m,b)$ \\
\midrule
\multicolumn{5}{@{}l}{\textit{Turner 2004}}\\
\midrule
\multirow{4}{*}{hc16 ligase} & $8$ & $(0.0,\,\phantom{-}0.0)$ & $(0.6,\,-0.4)$ & $(0.6,\,-0.4)$ \\
 & $17$ & $(2.0,\,\phantom{-}0.0)$ & $(1.2,\,\phantom{-}0.0)$ & $(1.8,\,-0.2)$ \\
 & $34$ & $(1.8,\,\phantom{-}0.0)$ & $(1.6,\,-0.2)$ & $(3.4,\,-0.6)$ \\
 & $57$ & $(2.2,\,-0.4)$ & $(1.4,\,-0.6)$ & $(3.6,\,-1.2)$ \\
\addlinespace
\multirow{4}{*}{\textit{Tetrahymena} ribozyme} & $8$ & $(1.4,\,\phantom{-}0.0)$ & $(0.4,\,\phantom{-}0.0)$ & $(1.6,\,-0.2)$ \\
 & $17$ & $(2.6,\,\phantom{-}0.0)$ & $(0.2,\,\phantom{-}0.0)$ & $(1.2,\,-0.2)$ \\
 & $34$ & $(4.4,\,-0.2)$ & $(0.2,\,\phantom{-}0.0)$ & $(1.8,\,-0.6)$ \\
 & $57$ & $(2.4,\,-0.4)$ & $(0.2,\,\phantom{-}0.0)$ & $(3.0,\,-0.8)$ \\
\addlinespace
\multirow{4}{*}{HCV IRES} & $8$ & $(0.6,\,-0.8)$ & $(0.0,\,-0.2)$ & $(0.2,\,\phantom{-}0.0)$ \\
 & $17$ & $(2.0,\,\phantom{-}0.0)$ & $(4.0,\,\phantom{-}0.0)$ & $(2.8,\,\phantom{-}0.0)$ \\
 & $34$ & $(4.4,\,-1.2)$ & $(5.0,\,\phantom{-}0.0)$ & $(4.6,\,-0.8)$ \\
 & $57$ & $(3.4,\,-1.4)$ & $(2.2,\,\phantom{-}0.0)$ & $(2.0,\,\phantom{-}0.0)$ \\
\addlinespace
\multirow{4}{*}{\textit{V.\,cholerae} gly.\ riboswitch} & $8$ & $(0.0,\,\phantom{-}0.0)$ & $(2.0,\,-0.4)$ & $(2.2,\,-0.6)$ \\
 & $17$ & $(2.6,\,\phantom{-}0.0)$ & $(2.4,\,-0.2)$ & $(3.6,\,-0.8)$ \\
 & $34$ & $(1.8,\,\phantom{-}0.0)$ & $(2.0,\,\phantom{-}0.0)$ & $(2.6,\,-3.0)$ \\
 & $57$ & $(2.8,\,-0.2)$ & $(2.2,\,-0.4)$ & $(2.0,\,-2.2)$ \\
\midrule
\multicolumn{5}{@{}l}{\textit{Andronescu 2007}}\\
\midrule
\multirow{4}{*}{hc16 ligase} & $8$ & $(1.4,\,-0.2)$ & $(1.8,\,-0.2)$ & $(0.8,\,-0.8)$ \\
 & $17$ & $(0.8,\,\phantom{-}0.0)$ & $(2.0,\,-0.2)$ & $(3.8,\,-0.4)$ \\
 & $34$ & $(1.4,\,\phantom{-}0.0)$ & $(3.6,\,-0.6)$ & $(4.0,\,-0.6)$ \\
 & $57$ & $(1.4,\,-0.6)$ & $(2.8,\,-1.0)$ & $(5.0,\,-1.6)$ \\
\addlinespace
\multirow{4}{*}{\textit{Tetrahymena} ribozyme} & $8$ & $(1.4,\,-0.2)$ & $(0.0,\,\phantom{-}0.0)$ & $(1.6,\,-0.4)$ \\
 & $17$ & $(1.4,\,\phantom{-}0.0)$ & $(0.0,\,\phantom{-}0.0)$ & $(0.0,\,-1.0)$ \\
 & $34$ & $(4.4,\,-0.6)$ & $(0.0,\,\phantom{-}0.0)$ & $(0.0,\,-1.0)$ \\
 & $57$ & $(2.4,\,-1.0)$ & $(0.0,\,\phantom{-}0.0)$ & $(4.8,\,-1.2)$ \\
\addlinespace
\multirow{4}{*}{HCV IRES} & $8$ & $(1.6,\,-0.4)$ & $(0.6,\,\phantom{-}0.0)$ & $(0.6,\,\phantom{-}0.0)$ \\
 & $17$ & $(0.6,\,\phantom{-}0.0)$ & $(0.6,\,\phantom{-}0.0)$ & $(0.6,\,\phantom{-}0.0)$ \\
 & $34$ & $(2.4,\,-0.6)$ & $(3.2,\,-0.2)$ & $(0.8,\,\phantom{-}0.0)$ \\
 & $57$ & $(3.6,\,-1.2)$ & $(1.0,\,\phantom{-}0.0)$ & $(0.8,\,\phantom{-}0.0)$ \\
\addlinespace
\multirow{4}{*}{\textit{V.\,cholerae} gly.\ riboswitch} & $8$ & $(0.2,\,-0.2)$ & $(1.0,\,\phantom{-}0.0)$ & $(0.0,\,\phantom{-}0.0)$ \\
 & $17$ & $(1.0,\,\phantom{-}0.0)$ & $(1.8,\,-0.2)$ & $(1.2,\,-0.8)$ \\
 & $34$ & $(4.0,\,-0.6)$ & $(1.6,\,\phantom{-}0.0)$ & $(1.6,\,-1.0)$ \\
 & $57$ & $(2.0,\,-1.0)$ & $(2.0,\,-0.6)$ & $(0.4,\,-0.2)$ \\
\bottomrule
\end{tabular}
\end{table*}

\subsection*{Combination of experiments performed in different settings: \textit{cspA} 5' UTR}\label{sec:cspA}

\begin{figure}[t]
\centering
\includegraphics[width=\linewidth]{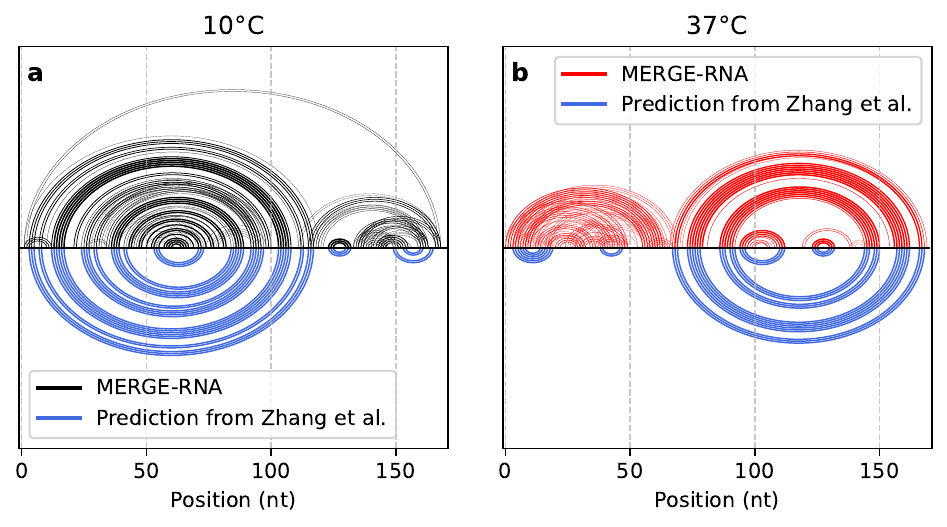}
\caption{Ensemble predictions for \textit{cspA} 5$^\prime$ UTR at two different temperatures.
\textbf{(a,b)} Arc plots corresponding to the DMS data collected at 10 and 37$^\circ$C, respectively.
Predictions from MERGE-RNA (red and black) are compared with
predictions from Ref.~\citep{zhangStressResponseThat2018} (blue).
}
\label{fig:cspA}
\end{figure}

\begin{figure}[h]
    \centering
    \includegraphics[width=.8\linewidth]{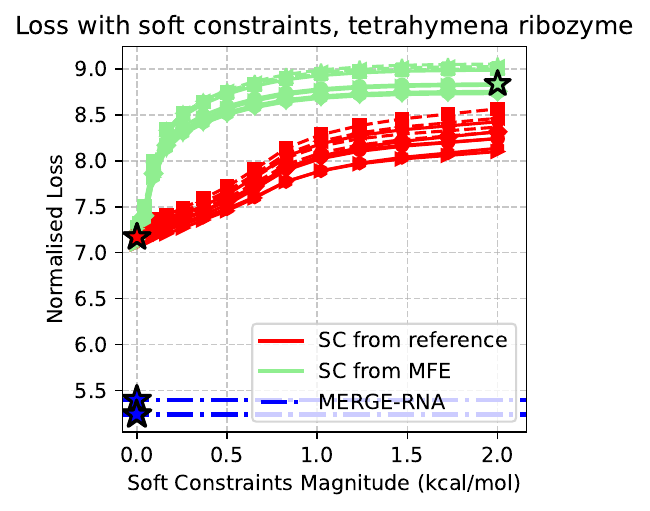}
    \caption{Same analysis as in Fig.~\ref{fig:fig2b_HCV_IRES}, for the \textit{Tetrahymena} ribozyme.}
    \label{fig:fig2b_tetrahymena_ribozyme}
\end{figure}

\begin{figure}[h]
    \centering
    \includegraphics[width=.8\linewidth]{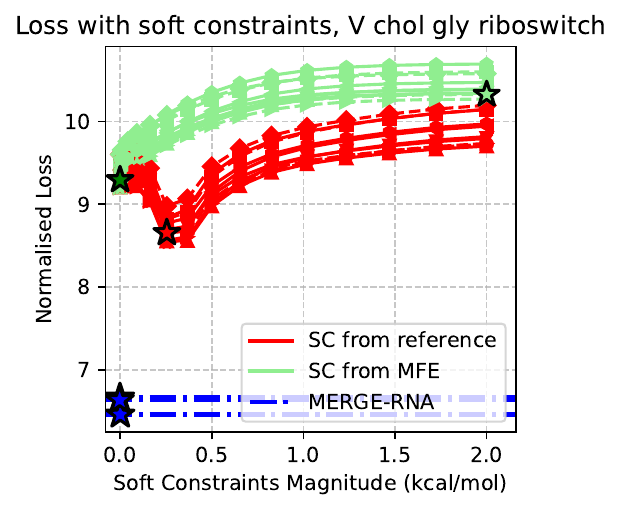}
    \caption{Same analysis as in Fig.~\ref{fig:fig2b_HCV_IRES}, for the \textit{V. cholerae} glycine riboswitch.}
    \label{fig:fig2b_V_chol_gly_riboswitch}
\end{figure}

\begin{figure}[h]
    \centering
    \includegraphics[width=.8\linewidth]{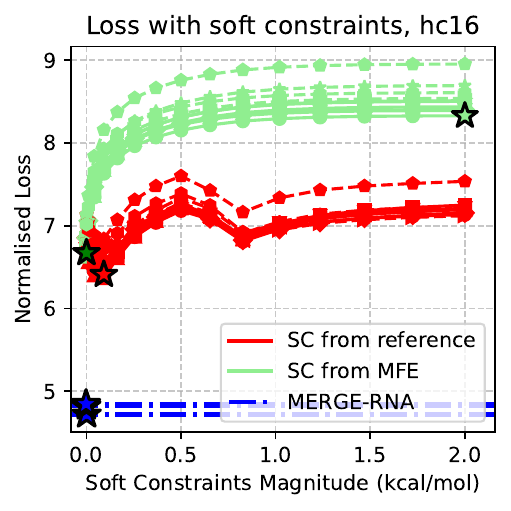}
    \caption{Same analysis as in Fig.~\ref{fig:fig2b_HCV_IRES}, for the hc16 ligase.}
    \label{fig:fig2b_hc16}
\end{figure}

\begin{figure}[h]
    \centering
    \includegraphics[width=\linewidth]{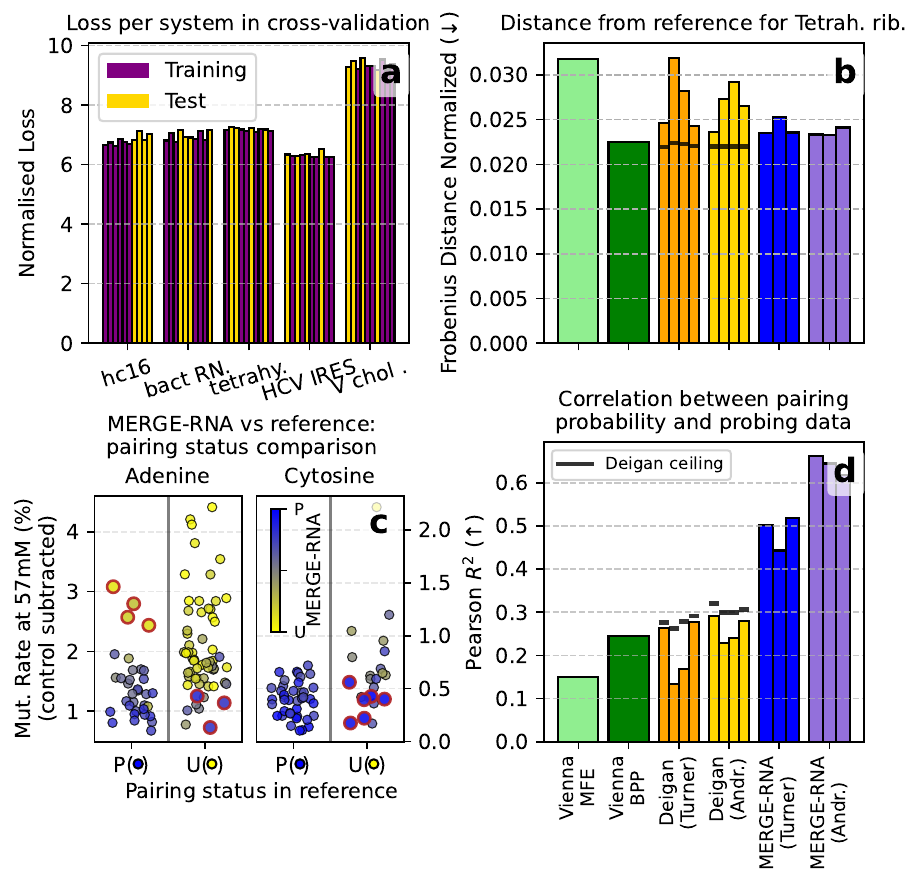}
    \caption{Analysis of \textit{Tetrahymena} ribozyme. This figure is analogous to Fig.~\ref{fig:redmond} in the main text, but here we show results for the \textit{Tetrahymena} ribozyme system.}
    \label{fig:fig2_tetrahymena_ribozyme}
\end{figure}

\begin{figure}[h]
    \centering
    \includegraphics[width=\linewidth]{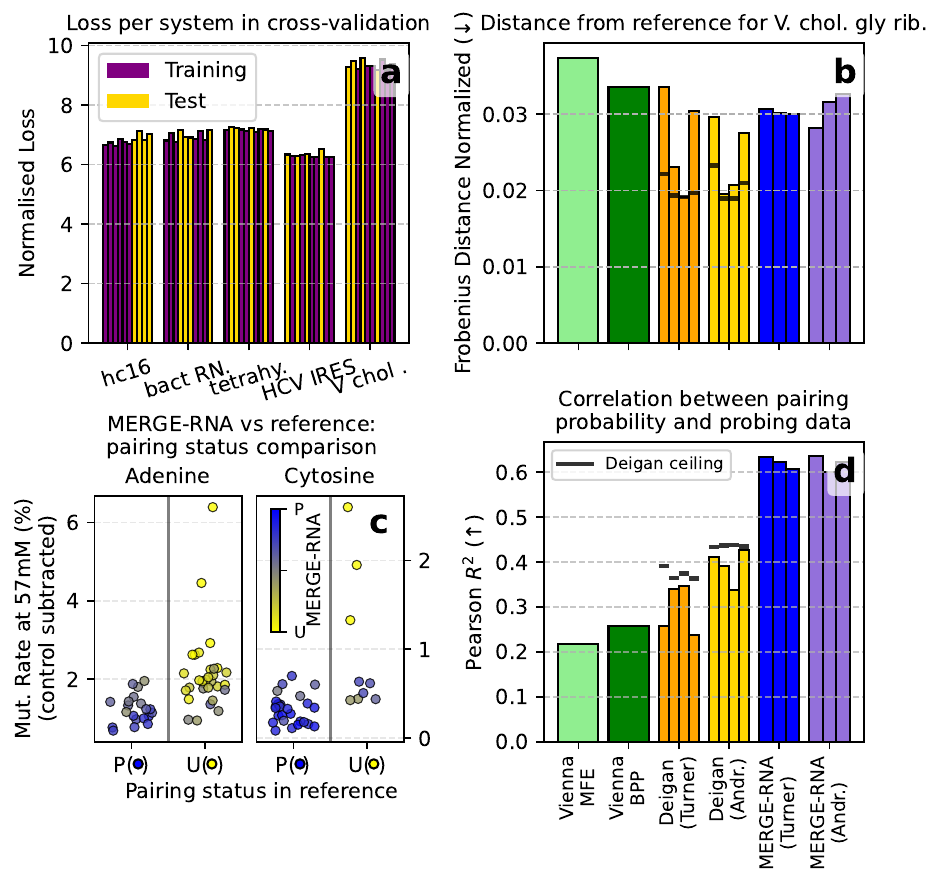}
    \caption{Analysis of \textit{V. cholerae} glycine riboswitch. This figure is analogous to Fig.~\ref{fig:redmond} in the main text, but here we show results for the \textit{V. cholerae} glycine riboswitch system.}
    \label{fig:fig2_V_chol_gly_riboswitch}
\end{figure}

\begin{figure}[h]
    \centering
    \includegraphics[width=\linewidth]{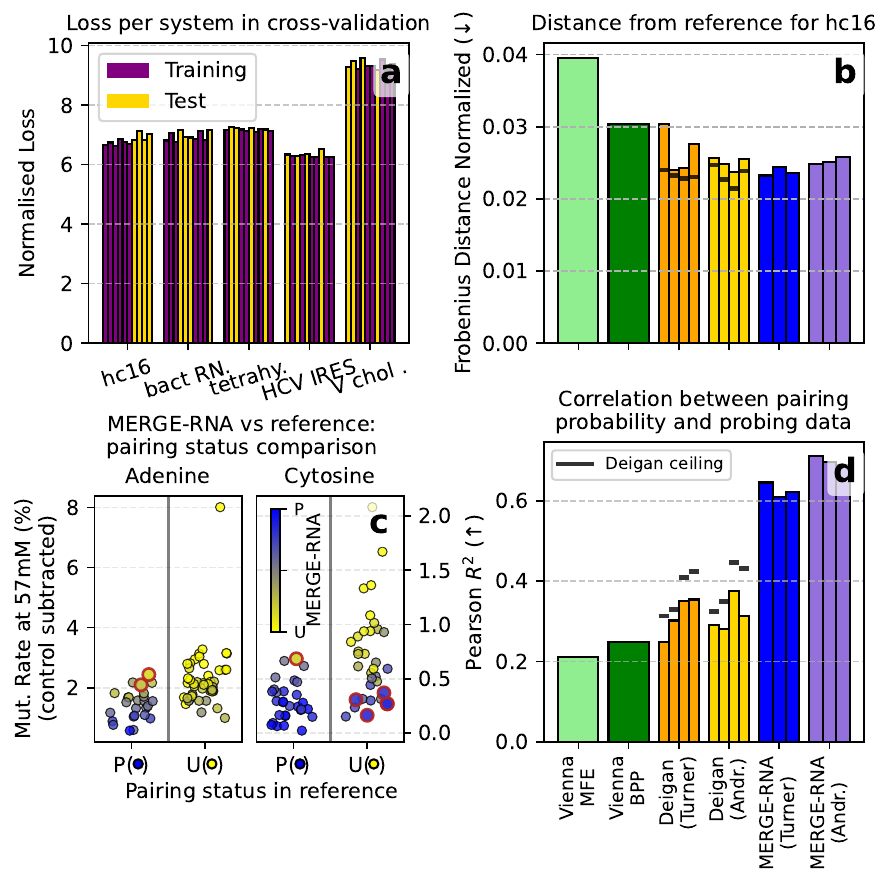}
    \caption{Analysis of HC16. This figure is analogous to Fig.~\ref{fig:redmond} in the main text, but here we show results for the HC16 system.}
    \label{fig:fig2_hc16}
\end{figure}

\begin{figure*}
    \centering
    \includegraphics[width=\linewidth]{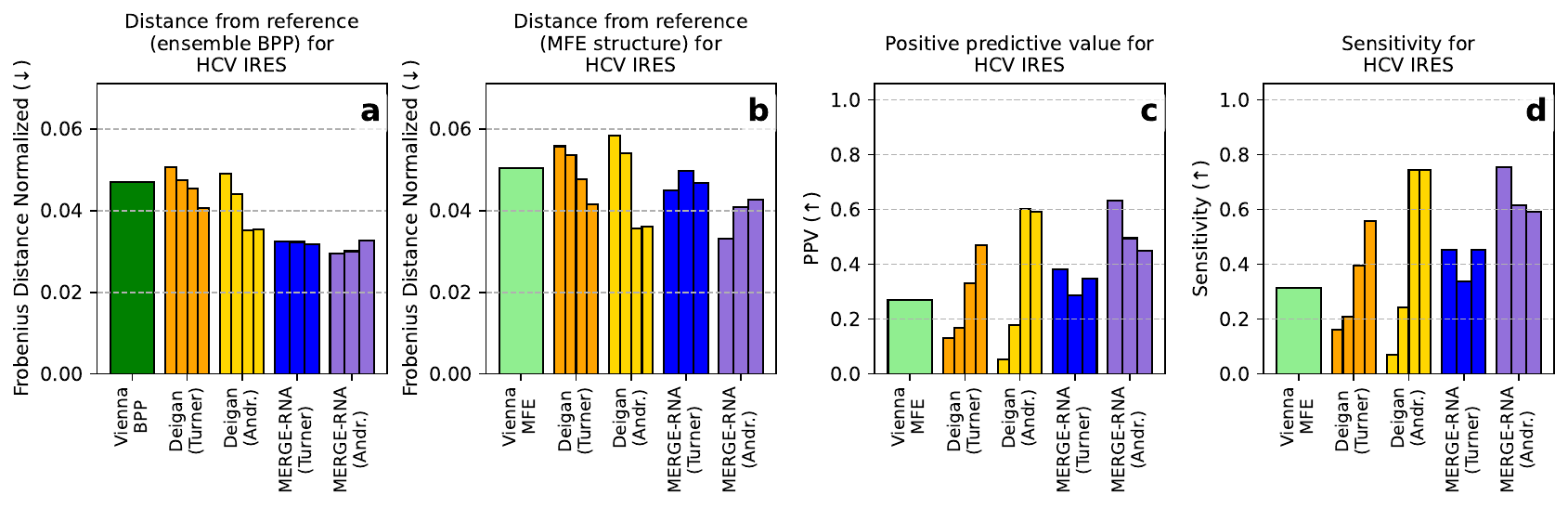}
    \caption{Structure-prediction metrics for the HCV IRES: no-data baseline (ViennaRNA), Deigan pseudo-free-energy approach and MERGE-RNA. \textbf{(a)} Frobenius distance to the reference computed on the partition-function ensemble of base-pairing probabilities, as in Fig.~\ref{fig:redmond}b; \textbf{(b)} the same distance computed on the single minimum-free-energy structure; \textbf{(c)} positive predictive value and \textbf{(d)} sensitivity of that structure, with $\pm1$-nt tolerance. Panels (a) and (b) share the vertical scale and differ only in the predicted object, isolating the effect of scoring a single MFE structure. Deigan sub-bars are the four DMS concentrations $[8, 17, 34, 57]$~mM with the leave-one-system-out coefficients; MERGE-RNA sub-bars are three independent fit runs.}
    \label{fig:deigan_metrics_HCV_IRES}
\end{figure*}

\begin{figure*}
    \centering
    \includegraphics[width=\linewidth]{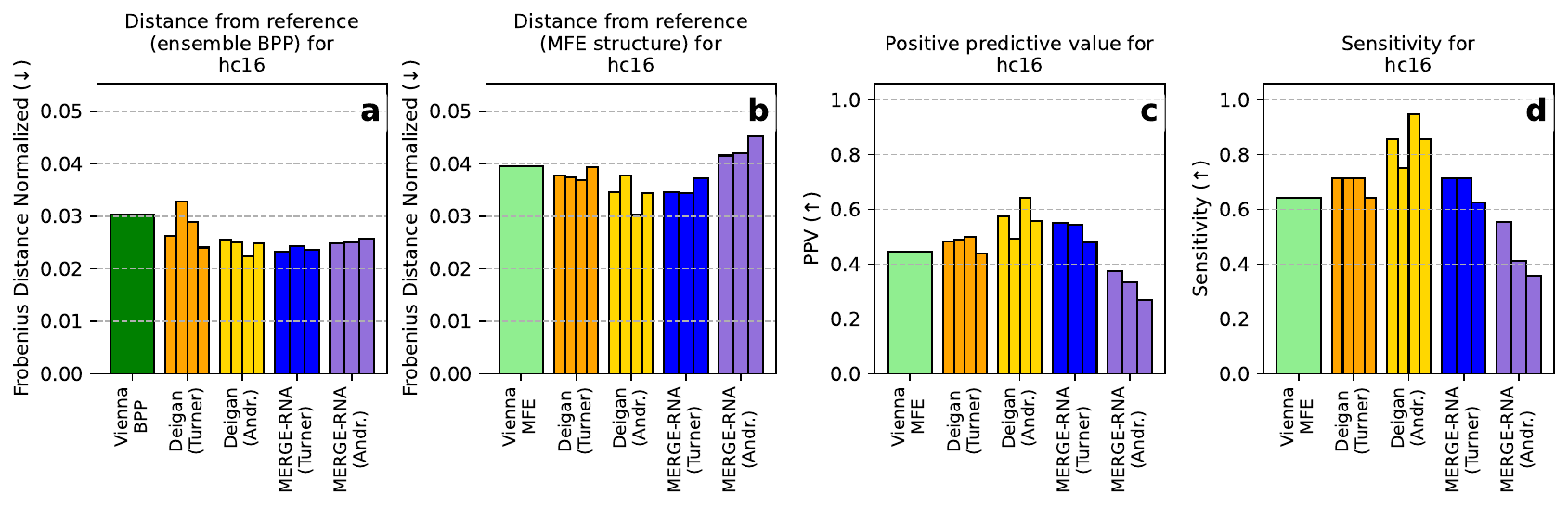}
    \caption{Structure-prediction metrics for the HC16 system, analogous to Fig.~\ref{fig:deigan_metrics_HCV_IRES}.}
    \label{fig:deigan_metrics_hc16}
\end{figure*}

\begin{figure*}
    \centering
    \includegraphics[width=\linewidth]{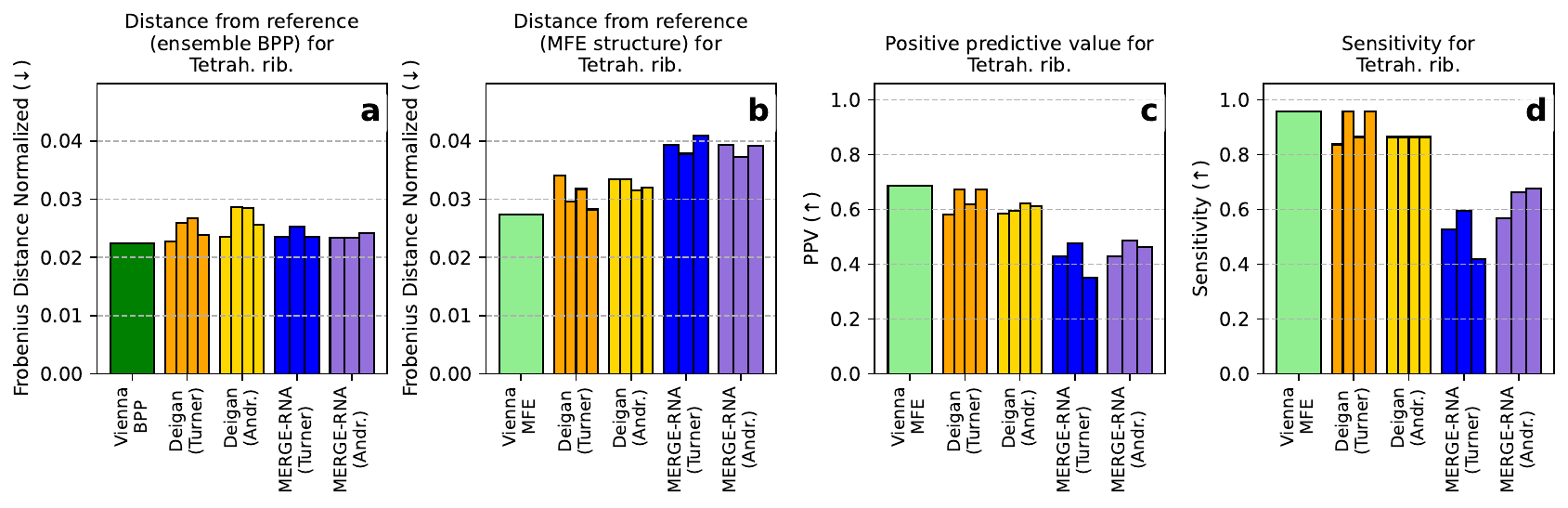}
    \caption{Structure-prediction metrics for the \textit{Tetrahymena} ribozyme, analogous to Fig.~\ref{fig:deigan_metrics_HCV_IRES}.}
    \label{fig:deigan_metrics_tetrahymena_ribozyme}
\end{figure*}

\begin{figure*}
    \centering
    \includegraphics[width=\linewidth]{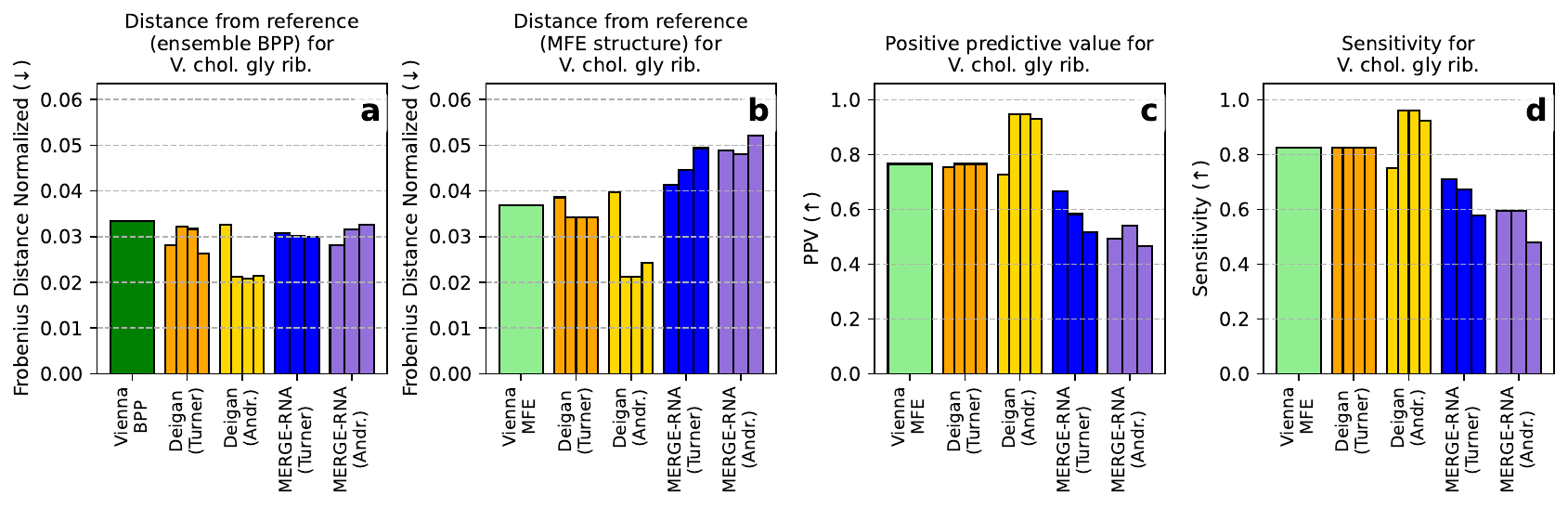}
    \caption{Structure-prediction metrics for the \textit{V. cholerae} glycine riboswitch, analogous to Fig.~\ref{fig:deigan_metrics_HCV_IRES}.}
    \label{fig:deigan_metrics_V_chol_gly_riboswitch}
\end{figure*}

\begin{figure*}
    \centering
    \includegraphics[width=.95\linewidth]{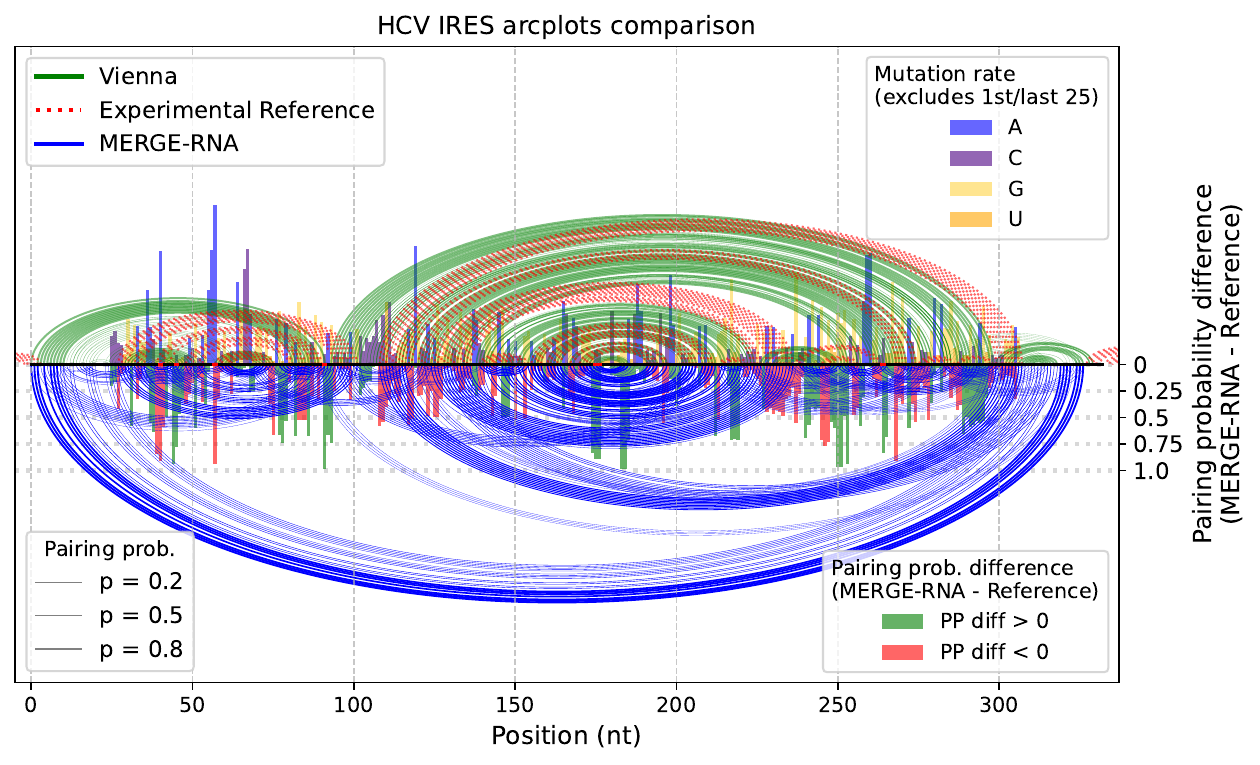}
    \caption{Detailed arc plot representation comparing the secondary structure of the reference structure (red arcs), the baseline thermodynamic ensemble (green arcs) and the MERGE-RNA refined ensemble (blue arcs) for the HCV IRES system. The arc thickness is proportional to the base-pairing probability. Bars above the horizontal line show the experimental mutation rate profiles at 57mM, colored by nucleotide: A (blue), C (violet), G (yellow) and U (orange). First and last 25 nucleotides are not shown since are not used in the model fitting.
    Bars below the horizontal line show the difference in base-pairing probability between MERGE-RNA and the binary reference structure, colored in green when positive (MERGE-RNA predicts more structure) and red otherwise.
    'Strong disagreements' as defined in the main text (where MERGE-RNA predictions are $<25\%$ or $>75\%$ while disagreeing with the reference structure for A and C nucleotides) are highlighted in red on the horizontal axis.}
    \label{fig:full_seq_arcplot_HCV_IRES_detailed}
\end{figure*}

\begin{figure*}
    \centering
    \includegraphics[width=.95\linewidth]{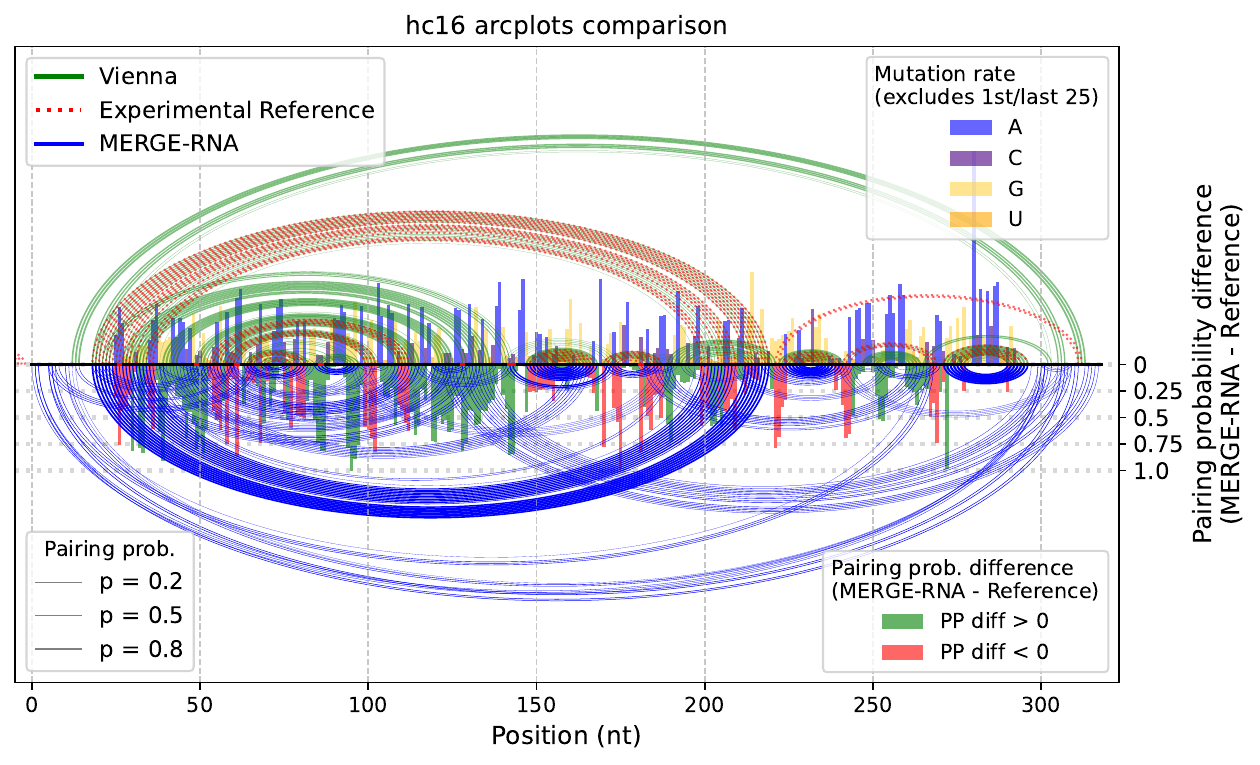}
    \caption{Detailed arc plot representation for the HC16 system, analogous to Fig.~\ref{fig:full_seq_arcplot_HCV_IRES_detailed}.}
    \label{fig:full_seq_arcplot_hc16_detailed}
\end{figure*}

\begin{figure*}
    \centering
    \includegraphics[width=\linewidth]{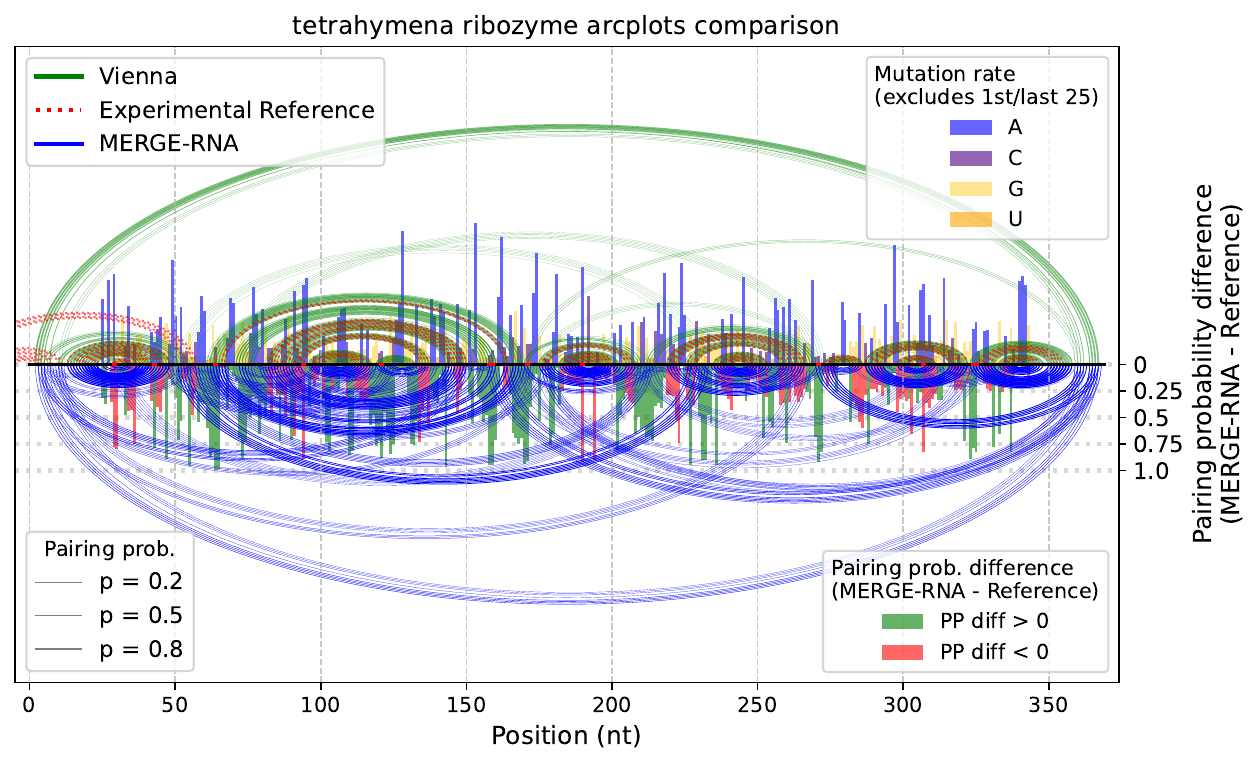}
    \caption{Detailed arc plot representation for the \textit{Tetrahymena} ribozyme system, analogous to Fig.~\ref{fig:full_seq_arcplot_HCV_IRES_detailed}.}
    \label{fig:full_seq_arcplot_tetrahymena_ribozyme_detailed}
\end{figure*}

\begin{figure*}
    \centering
    \includegraphics[width=\linewidth]{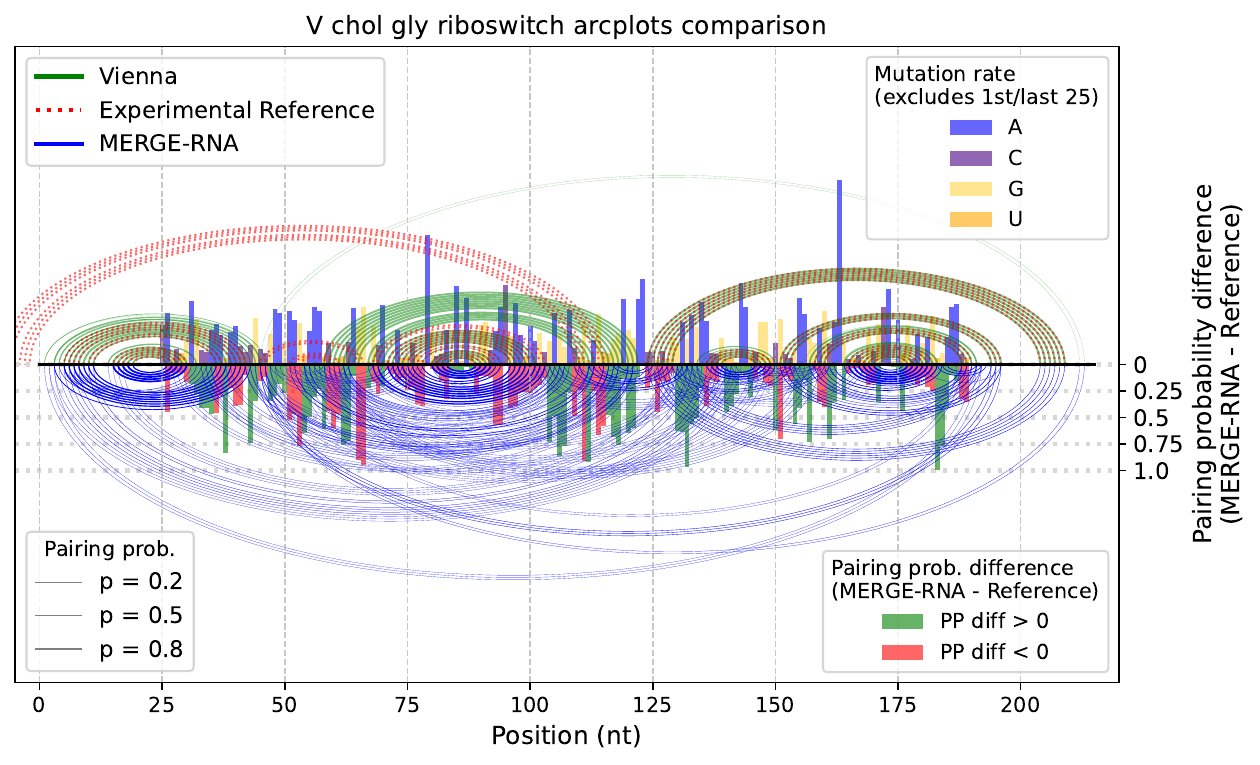}
    \caption{Detailed arc plot representation for the \textit{V. cholerae} glycine riboswitch system, analogous to Fig.~\ref{fig:full_seq_arcplot_HCV_IRES_detailed}.}
    \label{fig:full_seq_arcplot_V_chol_gly_riboswitch_detailed}
\end{figure*}

\begin{figure}
    \centering
    \includegraphics[width=.95\linewidth]{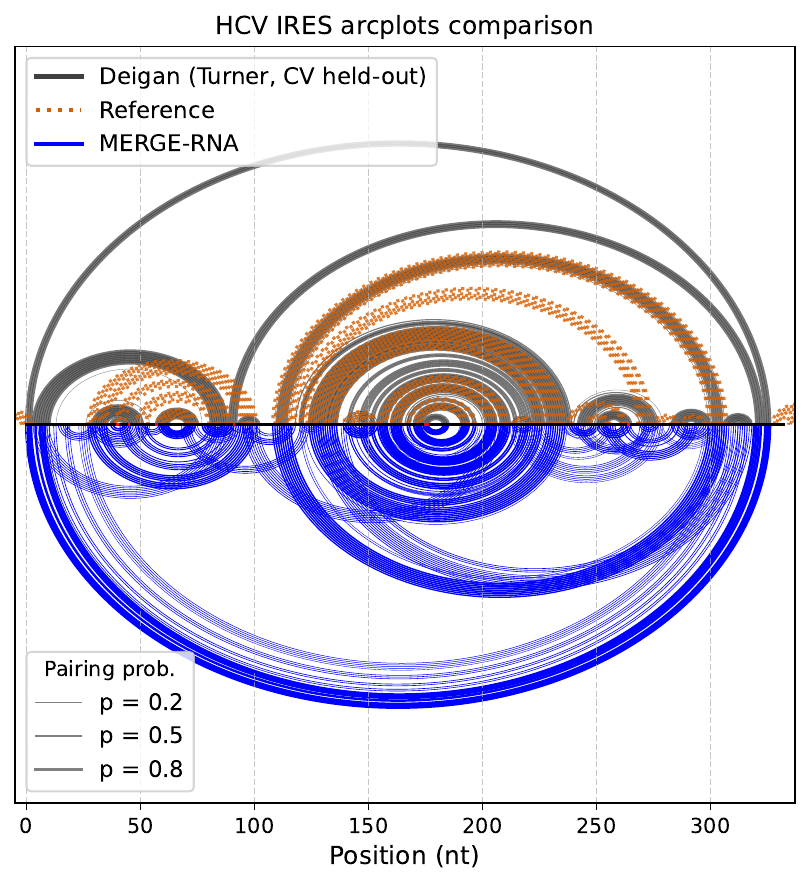}
    \caption{Arc plot for the HCV IRES, as Fig.~\ref{fig:arcplot} of the main text but with the upper solid arcs (dark grey) showing the Deigan ensemble instead of the unconstrained thermodynamic one, computed at $57$~mM under the Turner 2004 parameters with the leave-one-system-out coefficients of Table~\ref{tab:deigan_mb}. Reference (red dotted) and MERGE-RNA (blue) are unchanged, as are the strong-disagreement marks on the axis, which are computed between MERGE-RNA and the reference.}
    \label{fig:full_seq_arcplot_HCV_IRES_deigan}
\end{figure}

\begin{figure}
    \centering
    \includegraphics[width=.95\linewidth]{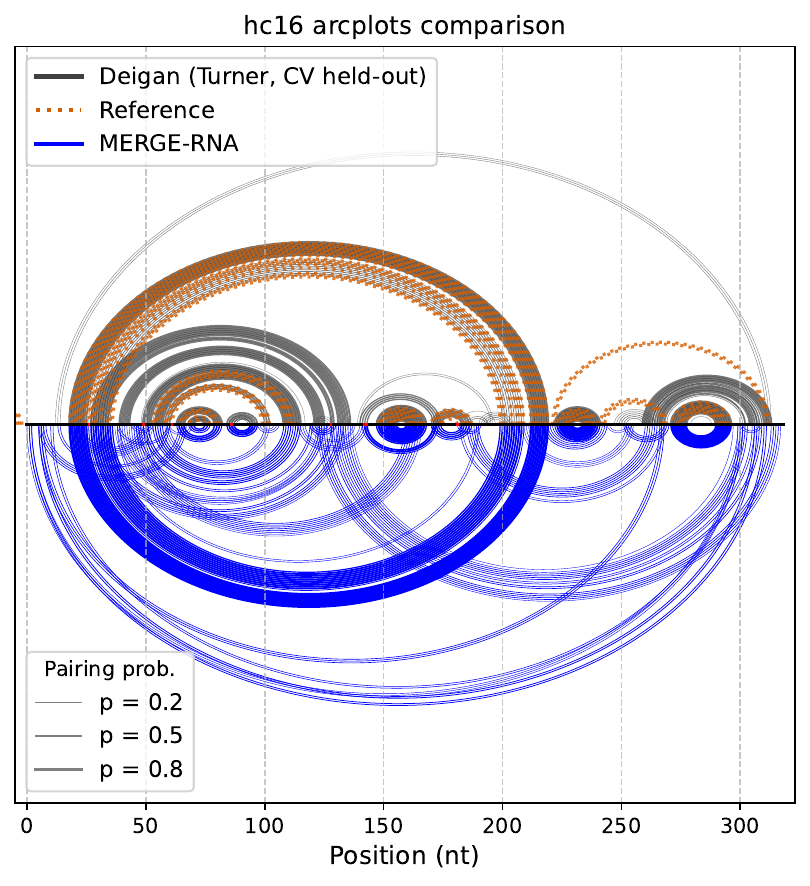}
    \caption{Arc plot representation with the Deigan ensemble for the HC16 system, analogous to Fig.~\ref{fig:full_seq_arcplot_HCV_IRES_deigan}.}
    \label{fig:full_seq_arcplot_hc16_deigan}
\end{figure}

\begin{figure}
    \centering
    \includegraphics[width=\linewidth]{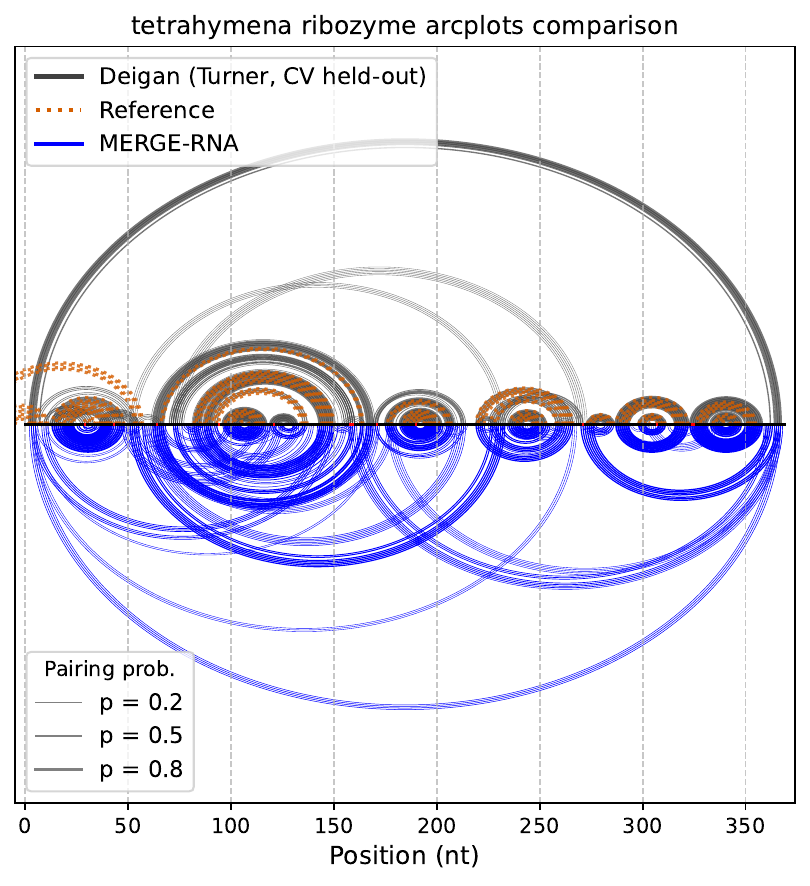}
    \caption{Arc plot representation with the Deigan ensemble for the \textit{Tetrahymena} ribozyme system, analogous to Fig.~\ref{fig:full_seq_arcplot_HCV_IRES_deigan}.}
    \label{fig:full_seq_arcplot_tetrahymena_ribozyme_deigan}
\end{figure}

\begin{figure}
    \centering
    \includegraphics[width=\linewidth]{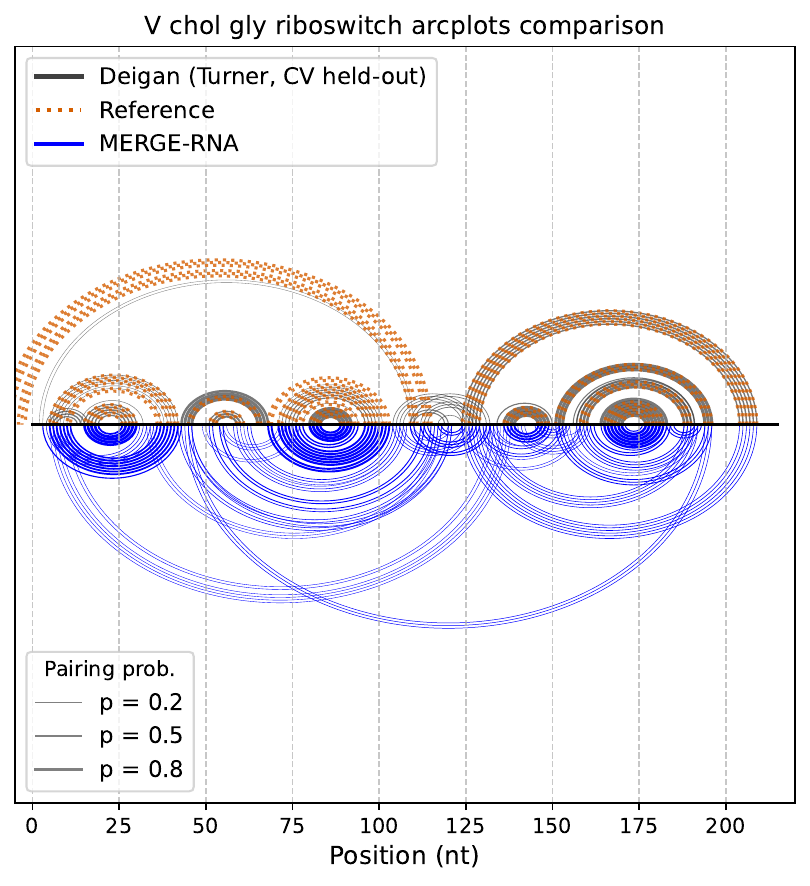}
    \caption{Arc plot representation with the Deigan ensemble for the \textit{V. cholerae} glycine riboswitch system, analogous to Fig.~\ref{fig:full_seq_arcplot_HCV_IRES_deigan}.}
    \label{fig:full_seq_arcplot_V_chol_gly_riboswitch_deigan}
\end{figure}

\begin{figure*}
    \centering
    \includegraphics[width=\linewidth]{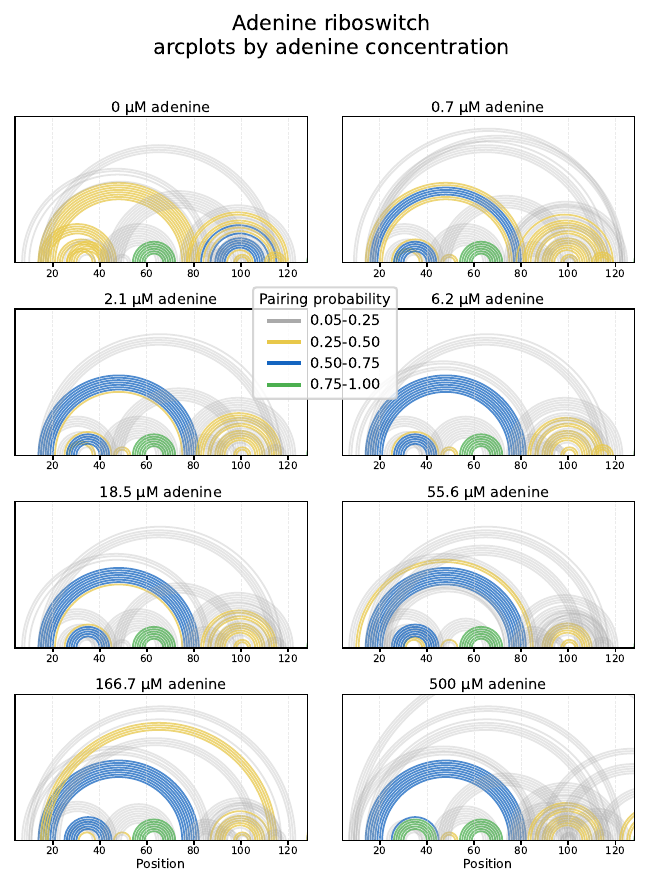}
    \caption{Arc plot representation for the MERGE-RNA inferred pairing probabilities for the \textit{V. vulnificus} adenine riboswitch, when trained on data collected at different ligand concentrations. Colors indicate the base-pairing probability.}
    \label{fig:adenine_riboswitch_arcgrid}
\end{figure*}

\begin{figure}[!tbp]
\centering
\includegraphics[width=\linewidth]{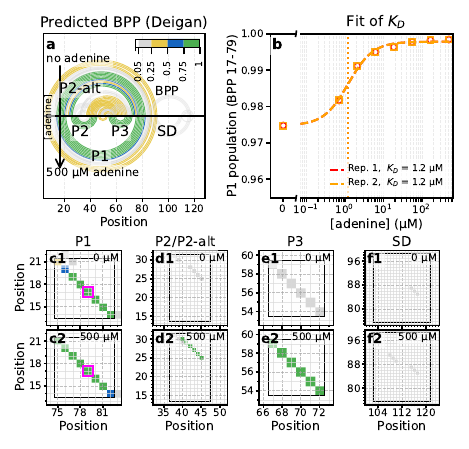}
\caption{Same analysis as in Fig.~\ref{fig:adenine_riboswitch_maintext} of the main text, for the \textit{V. vulnificus} adenine riboswitch, but obtained with the Deigan pseudo-free-energy approach (\nameref{si:deigan_refit}) instead of MERGE-RNA.}
\label{fig:adenine_riboswitch_maintext_deigan}
\end{figure}

\begin{figure*}
    \centering
    \includegraphics[width=\linewidth]{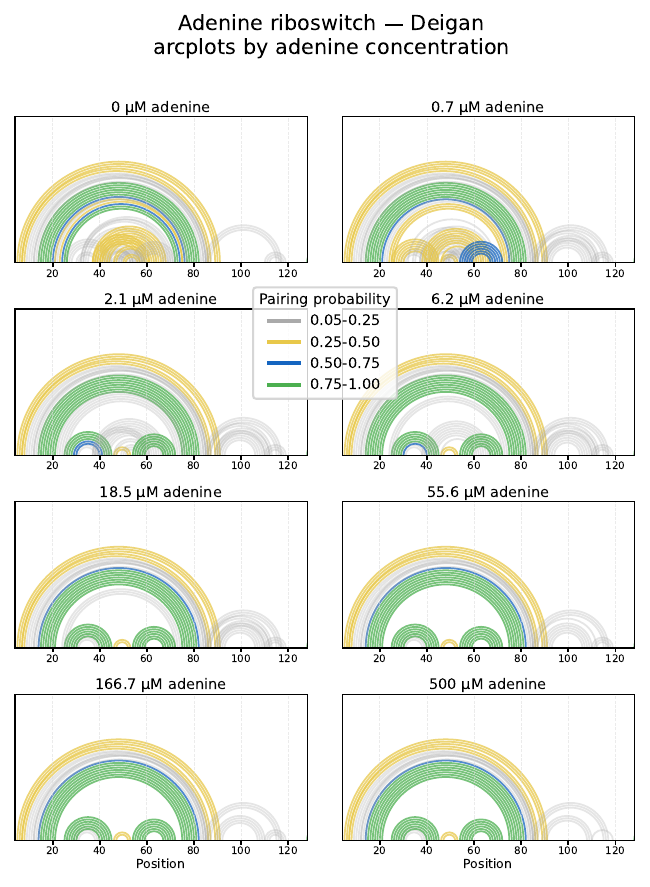}
    \caption{Same analysis as in Fig.~\ref{fig:adenine_riboswitch_arcgrid}, but obtained with the Deigan pseudo-free-energy approach. Colors indicate the base-pairing probability.}
    \label{fig:adenine_riboswitch_arcgrid_deigan}
\end{figure*}

We here test the capability of MERGE-RNA to operate in situations where experiments performed in different settings are combined, namely in
capturing complex structural rearrangements by
considering two DMS mutational profiles for \textit{cspA} 5$^\prime$ UTR recorded
at 10 and 37$^\circ$C \citep{zhangStressResponseThat2018}.
This sequence acts as a thermoswitch \citep{giuliodoriCspAMRNAThermosensor2010},
with different structures at the two reported temperatures.
The two experiments have been recorded at different probing times (2 hours vs 10 minutes).
Our approach has been designed to analyze homogeneous experiments and
does not model the dependence of the observed mutation rate
on the probing time.
We hence normalize the dataset at 10$^\circ$C to the same average mutation rate observed
at 37$^\circ$C. In addition, to avoid confusing the effect of the mutational profiles with the
temperature-dependence of the thermodynamic model, we use the thermodynamic model set on the same intermediate temperature
for every ensemble (23.5$^\circ$C).
Finally, we note that these experimental datasets lack a control mutation profile obtained in absence of DMS,
making the separation of the physical signal from the systematic bias more difficult for our model.
Given the mentioned limitations, we decided to perform a set of 6 independent fits, starting from different random initializations of the physical parameters.
We report in Fig.~\ref{fig:cspA} the ensemble predicted by the model that achieved the lowest loss,
while additional fits from independent random initializations are shown in Figs~\siref{fig:fig:fig4_cspA_fix_1_37_10}--\siref{fig:fig:fig4_cspA_fix_5_37_10}.

In Fig.~\ref{fig:cspA}(a,b), the structural ensembles predicted by MERGE-RNA largely recapitulate the individual structures reported by Zhang et al.\ \citep{zhangStressResponseThat2018}.
At 10$^\circ$C and 37$^\circ$C, our ensembles closely match their predictions in positions 1--120 and 70--170, respectively, but exhibit notable differences in other regions, particularly in segments where alternative helices compete.
Unlike Ref.~\citep{zhangStressResponseThat2018}, which reports a single minimum-free-energy structure, MERGE-RNA predicts a thermodynamic ensemble that accounts for multiple competitive helices and their relative populations.
Overall, this analysis demonstrates that our model can capture complex structural rearrangements in biological systems.
For comparison, the corresponding predictions obtained with the optimized Deigan pseudo-free-energy approach are shown in Fig.~\ref{fig:cspA_deigan}.

\clearpage

\section*{Supplementary Figures and Tables}

\begin{sidewaystable}[h]
\centering
\begin{tabular}{l|*{7}{r@{.}l}}
\toprule
Training sets &
\multicolumn{2}{c}{$\mu_r$(kcal/mol)} &
\multicolumn{2}{c}{$\Delta\mu_{\text{pairing}}$ (kcal/mol)} &
\multicolumn{2}{c}{$p_{\text{bind}}(\text{A},\text{unpaired})$} &
\multicolumn{2}{c}{$p_{\text{bind}}(\text{C},\text{unpaired})$} &
\multicolumn{2}{c}{$p_{\text{bind}}(\text{G})$} &
\multicolumn{2}{c}{$p_{\text{bind}}(\text{U})$} &
\multicolumn{2}{c}{$m_1$} \\
\midrule
hc16 + HCV + V.chol. & -0 & 08 & 0 & 34 & 1 & 00 & 0 & 60 & 0 & 48 & 0 & 05 & 0 & 53 \\
hc16 + RNaseP + HCV  & -0 & 05 & 0 & 49 & 1 & 00 & 0 & 57 & 0 & 57 & 0 & 05 & 0 & 50 \\
hc16 + RNaseP + Tetrah. & -0 & 02 & 0 & 15 & 0 & 89 & 0 & 56 & 0 & 28 & 0 & 02 & 0 & 56 \\
hc16 + RNaseP + V.chol. & -0 & 36 & 0 & 03 & 1 & 00 & 0 & 66 & 0 & 31 & 0 & 04 & 1 & 05 \\
hc16 + Tetrah. + HCV & -0 & 06 & 0 & 61 & 0 & 98 & 0 & 44 & 0 & 58 & 0 & 03 & 0 & 50 \\
hc16 + Tetrah. + V.chol. & -0 & 07 & 0 & 40 & 0 & 91 & 0 & 64 & 0 & 40 & 0 & 03 & 0 & 53 \\
RNaseP + HCV + V.chol. & 0 & 08 & 0 & 18 & 0 & 99 & 0 & 58 & 0 & 35 & 0 & 05 & 0 & 39 \\
RNaseP + Tetrah. + HCV & -0 & 08 & 0 & 24 & 0 & 97 & 0 & 51 & 0 & 35 & 0 & 04 & 0 & 53 \\
RNaseP + Tetrah. + V.chol. & -0 & 05 & 0 & 79 & 0 & 91 & 0 & 61 & 0 & 41 & 0 & 07 & 0 & 54 \\
Tetrah. + HCV + V.chol. & -0 & 09 & 0 & 50 & 1 & 00 & 0 & 49 & 0 & 57 & 0 & 05 & 0 & 45 \\
\bottomrule
\end{tabular}
\caption{Physical parameters obtained by training MERGE-RNA on different triplets of RNA systems (see section 'Transferability and accuracy of the physical model' in the main text).}
\label{tab:physical_params}
\end{sidewaystable}


\newpage

\clearpage

\clearpage

\clearpage

\clearpage

\foreach \i in {1,...,5}{
    \begin{figure}[h]
        \centering
        \includegraphics[width=\linewidth]{figs/fig4_cspA_fix_\i_37_10.pdf}
        \caption{Arc plot visualization of the \textit{cspA} $5^\prime$ UTR showing MERGE-RNA ensembles at 10$^\circ$C (black) and 37$^\circ$C (red), compared to the single structure prediction from ref.~\citep{zhangStressResponseThat2018}. This figure is analogous to Fig.~\ref{fig:cspA}, but shows results from a different random initialization of the fitting parameters (the main text presents the solution with lowest loss).}
        \label{fig:fig:fig4_cspA_fix_\i_37_10}
    \end{figure}
}

\begin{figure}[!tbp]
\centering
\includegraphics[width=\linewidth]{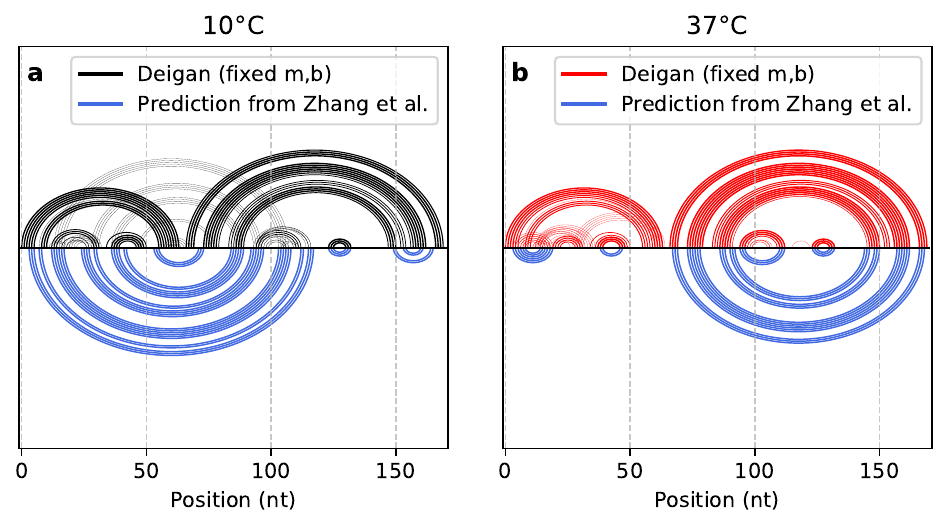}
\caption{Same analysis as in Fig.~\ref{fig:cspA}, but obtained with the Deigan pseudo-free-energy approach (fixed $m,b$) instead of MERGE-RNA.
\textbf{(a,b)} Arc plots for the DMS data collected at 10 and 37$^\circ$C, respectively.
The Deigan predictions (black and red) are compared with the single-structure prediction of Ref.~\citep{zhangStressResponseThat2018} (blue).}
\label{fig:cspA_deigan}
\end{figure}

\newpage
\begin{table}[h]
\centering
\caption{Loop co-occupancy inferred from baseline thermodynamic ensembles and after MERGE-RNA refinement for the putative bistable sequence (see section 'a putatively bistable sequence exhibits a heterogeneous ensemble with strand displacement'). Entries report the percentage of structures in which Loop1 and Loop2 are formed, using either the Turner2004
\cite{mathewsIncorporatingChemicalModification2004} or Andronescu2007
\cite{andronescuEfficientParameterEstimation2007} baseline parameter sets.}
\label{tab:loop_cooccupancy}
\begin{tabular}{|l|c|c|}
\hline
 & Baseline & MERGE-RNA \\
\hline
Turner & \begin{tabular}{c|cc}
\multicolumn{1}{c}{} & \multicolumn{2}{c}{Loop1} \\
Loop2 & 0 & 1 \\
\hline
0 & 0\% & 49\% \\
1 & 51\% & 0\% \\
\end{tabular} & \begin{tabular}{c|cc}
\multicolumn{1}{c}{} & \multicolumn{2}{c}{Loop1} \\
Loop2 & 0 & 1 \\
\hline
0 & 0\% & 27\% \\
1 & 16\% & 57\% \\
\end{tabular} \\
\hline
Andr. & \begin{tabular}{c|cc}
\multicolumn{1}{c}{} & \multicolumn{2}{c}{Loop1} \\
Loop2 & 0 & 1 \\
\hline
0 & 0\% & 19\% \\
1 & 81\% & 0\% \\
\end{tabular} & \begin{tabular}{c|cc}
\multicolumn{1}{c}{} & \multicolumn{2}{c}{Loop1} \\
Loop2 & 0 & 1 \\
\hline
0 & 0\% & 42\% \\
1 & 16\% & 42\% \\
\end{tabular} \\
\hline
\end{tabular}
\end{table}

\begin{figure*}
    \centering
    \includegraphics[width=\linewidth]{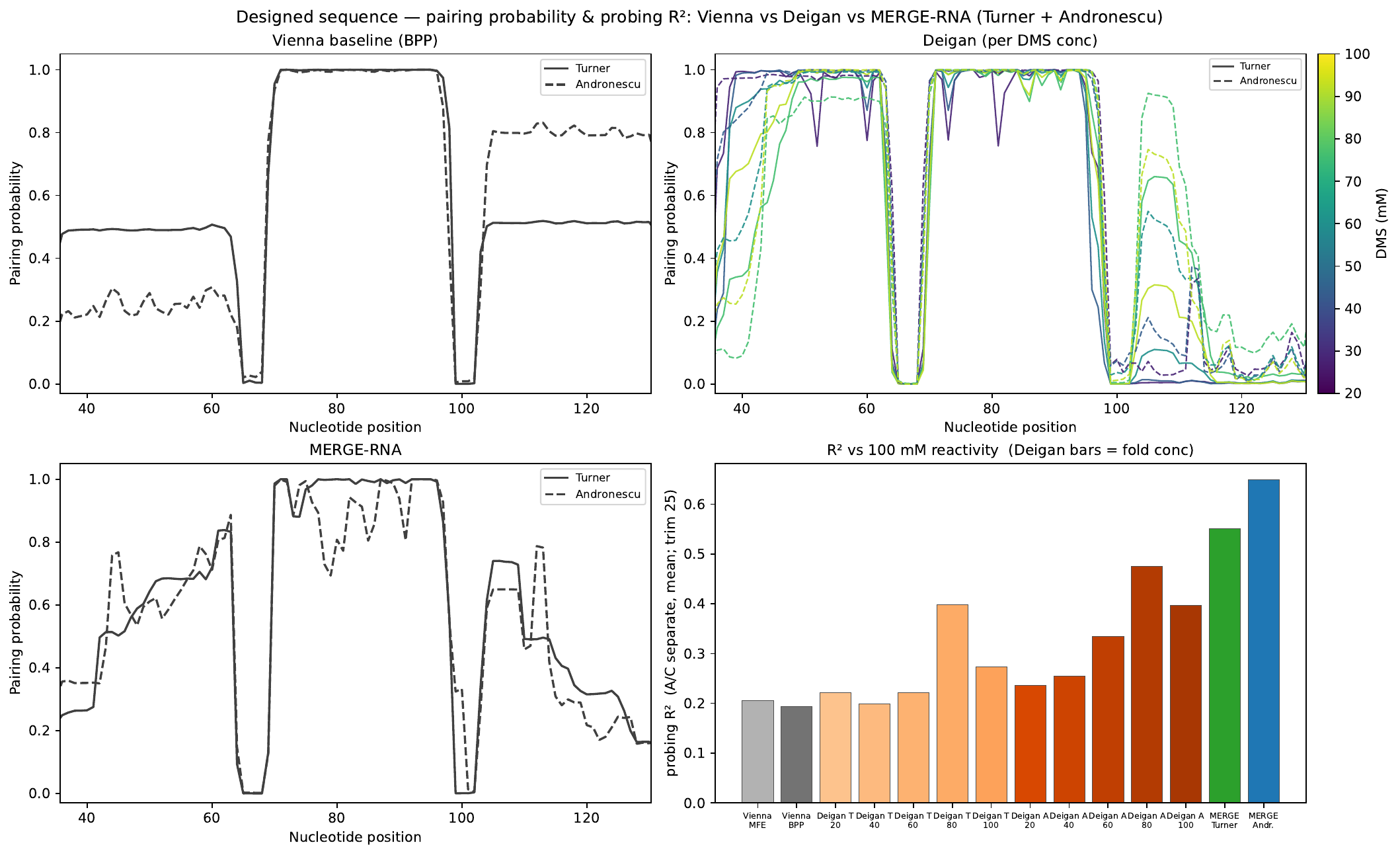}
    \caption{Same analysis as in Fig.~\ref{fig:fig_newseq}, for the designed sequence, comparing the Vienna baseline, the Deigan pseudo-free-energy approach (per DMS concentration), and MERGE-RNA under the Turner 2004 and Andronescu 2007 parameters. The bottom-right panel reports the probing $R^2$ against the 100~mM reactivity profile.}
    \label{fig:designed_deigan_2x2}
\end{figure*}

\end{document}